\documentclass[11pt]{article}

\usepackage{fullpage,amssymb,graphicx,amsfonts,amsmath,mathrsfs}
\usepackage{setspace}
\usepackage{natbib}
\usepackage[]{color}

\begin{document}
\title{Bayesian nonparametric mean residual life regression: Supplementary Material}

\author{VALERIE POYNOR \\
\textit{Department of Mathematics, California State University, Fullerton} \\[2pt]
ATHANASIOS KOTTAS$^\ast$\\
\textit{Department of Applied Mathematics and Statistics,
University of California, Santa Cruz}
\\[1pt]
{thanos@soe.ucsc.edu}}

\markboth%
{V. Poynor and A. Kottas}
{Nonparametric Bayesian inference for mean residual life functions}
\date{}
\vspace{-2pt}

\maketitle

\begin{center} 
{Abstract}
\end{center}
The mean residual life function is a key functional for a survival
distribution. It has a practically useful interpretation as the
expected remaining lifetime given survival up to a particular time
point, and it also characterizes the survival distribution. However,
it has received limited attention in terms of inference methods under
a probabilistic modeling framework. We seek to provide general
inference methodology for mean residual life regression. Survival data
often include a set of predictor variables for the survival response
distribution, and in many cases it is natural to include the
covariates as random variables into the modeling. We thus employ
Dirichlet process mixture modeling for the joint stochastic mechanism
of the covariates and survival responses. This approach implies a
flexible model structure for the mean residual life of the conditional
response distribution, allowing general shapes for mean residual life
as a function of covariates given a specific time point, as well as a
function of time given particular values of the covariate vector. To
expand the scope of the modeling framework, we extend the mixture
model to incorporate dependence across experimental groups, such as
treatment and control groups. This extension is built from a dependent
Dirichlet process prior for the group-specific mixing distributions,
with common locations and weights that vary across groups through
latent bivariate Beta distributed random variables. We develop
properties of the regression models, and discuss methods for prior
specification and posterior inference. The different components of the
methodology are illustrated with simulated data examples, and the
model is also applied to a data set comprising right censored survival
times.

\newpage
 
\section{Introduction}

The mean residual life (MRL) function of a continuous positive-valued random variable, $T$, 
provides the expected remaining lifetime given survival up to time $t$, $m( t) =$
$\text{E}(T - t \mid T>t)$. Its definition requires that $T$ has finite mean, which is given 
by $\text{E}(T) = m(0)$. The MRL function can be defined through the survival function, 
$S(t)=$ $\text{Pr}(T>t)$, in particular, $m(t) =$ 
$\left[\int_t^\infty S(u) \text{d}u\right]/S(t)$, with $m(t) \equiv 0$ when $S(t) = 0$.
Conversely, the survival function is defined through the MRL function via $S(t)=$ 
$\{ m(0)/m(t) \} \exp[ -\int_{0}^{t} \{ 1/m(u) \} \text{d}u]$ \citep{hall:wellner81}, 
and thus the MRL function characterizes the survival distribution. Given this property and its useful 
interpretation, the MRL function is of practical importance in a variety of fields, such as reliability, 
medicine, and actuarial science. 
%

Often associated with the survival times is a set of covariates, ${\boldsymbol x}$. 
The MRL regression function at a specified set of covariate values is given by:
\begin{eqnarray} 
m( t \mid {\boldsymbol x}) = \text{E}(T - t \mid T>t, {\boldsymbol x}) = 
\frac{\int_t^\infty S(u \mid {\boldsymbol x}) \text{d}u}{S(t \mid {\boldsymbol x})}  \label{eqn:mrl}
\end{eqnarray}
provided $\text{E}(T \mid {\boldsymbol x})<\infty$.
In the regression setting, it is of interest to develop modeling that allows flexible MRL function 
shapes over the covariate space. Note that MRL regression goes beyond standard regression problems
in that it involves study of covariate effects on a function over time. It is of interest to study how MRL
regression relationships change across different time points (mean regression corresponds to $t=0$),
as well as how the MRL function evolves across different covariate values. To our knowledge, there is 
no work in the Bayes nonparametrics literature that explores modeling and inference for the MRL 
function in the presence of covariates. 
%

Classical estimation methods for MRL regression have been primarily derived from the proportional MRL 
model, $m(t) =$ $\gamma \, m_0(t)$, where $\gamma >0$ and $m_0(t)$ is a baseline MRL function \citep{oakes:dasu}. If $\gamma <1$, the survival function, $S(t)$, associated with $m(t)$ is proper for any 
proper MRL function $m_0(t)$. Alternatively, $S(t)$ is proper for all $\gamma > 0$ if and only if $m_0(t)$ is nondecreasing. \cite{maguluri} extend the proportional MRL model to incorporate covariates, such that the 
MRL regression function is given by $m_0(t) \exp({\boldsymbol \beta}^T{\boldsymbol x})$, 
where ${\boldsymbol \beta}$ are the regression coefficients and $m_0(t)$ is the unknown baseline MRL function.  They propose two estimators for ${\boldsymbol \beta}$ in the case of fully observed survival responses.  
\cite{chen:cheng} expand the estimation methods under the semiparametric proportional MRL model 
to include censored responses, \cite{chen-wang} address the case with the censoring indicators missing 
at random, and \cite{bai-huang-zhou} account for right censored length-biased data.
Alternative to the proportional MRL model structure, \cite{chen06} and \cite{chen:cheng06} 
develop a class of additive MRL models, under which the MRL regression function is given by 
$m_0(t) + {\boldsymbol \beta}^T{\boldsymbol x}$.  
\cite{sun:zhang} generalize the class of additive MRL models by applying a pre-specified transformation $g$
to the regression function, where $g$ must be such that $g(m_0(t) + {\boldsymbol \beta}^T{\boldsymbol x})$ 
is a proper MRL function for all ${\boldsymbol x}$. 
%
%
This model is further extended in \cite{sun:song:zhang} to incorporate time-dependent regression parameters 
in a linear fashion. While these methods are more general than the basic approach of linking covariates through parameters of a fully parametric MRL function, they are still restricted by the proportional or additive form of 
the MRL regression function, by the parametric introduction of covariate effects, and by the fact that inference 
is not based on a fully probabilistic model setting.

Our objective is to develop a modeling framework that lends full and flexible inference for 
MRL regression within and across the covariate space. To accommodate the regression setting,
we extend our earlier work on inference for MRL functions, based on Dirichlet process (DP) 
mixture priors for the survival distribution \citep{poynor:kottas}.
To this end, we propose nonparametric DP mixture modeling for the joint stochastic
mechanism of the covariates and survival responses, from which inference for MRL regression emerges
through the implied conditional response distribution. This DP mixture density regression approach was 
proposed by \cite{muller:erk:west} for real-valued responses, and has been more recently 
elaborated under different settings; see, e.g., \cite{mullerquintana}, \cite{taddy:kottas}, 
\cite{wadeetal}, \cite{papa:richardson:best}, and \cite{deyoreokottas}.
For problems with a small to moderate number of random covariates, this modeling 
approach is attractive in terms of its inferential flexibility. At the same time, survival data typically 
comprise responses (and associated covariates) from subjects assigned to different experimental groups, 
such as control and treatment groups. The treatment indicator can not be meaningfully incorporated 
into the joint response-covariate mixture model as an additional component of the 
mixture kernel. We thus extend the model to allow distinct mixing distributions for the different 
groups, which are however dependent in the prior with the dependence built in a nonparametric 
fashion. We develop this extension in the context of two groups, using a dependent Dirichlet 
process (DDP) prior for the group-specific mixing distributions. A key aspect of the modeling approach 
is the choice of the mixture kernel that corresponds to the survival responses. Moreover, even though 
we do not model directly the MRL function of the response distribution, the implied 
model for the MRL function given the covariates has an appealing structure as a locally weighted 
mixture of the kernel MRL functions, with weights that depend on both time and the covariates.

The outline of the paper is as follows. Section~\ref{sec:curve_reg} develops the approach to modeling
and inference for MRL regression, including illustrations with synthetic data sets. In Section~\ref{sec:ddp}, 
we present the model elaboration to incorporate survival data from different experimental groups. We study properties of the proposed DDP prior model (with technical details included in Appendix A), and present 
results from two simulation data examples. In Section~\ref{sec:lung_data}, we provide a detailed analysis 
of a standard data set from the literature on right censored survival
times for patients with small cell lung cancer. 
Finally, Section~\ref{sec:conc} concludes with a summary.
%

\section{Mean residual life regression}
\label{sec:curve_reg}

\subsection{Model formulation}
\label{DPM_for_MRL}

For survival regression problems with a small/moderate number of random covariates, 
it is meaningful to model the joint distribution of covariates and survival responses.
A key benefit of this modeling approach for MRL regression revolves around the 
interpretable implied form for the MRL function of the conditional response distribution, 
which allows for general shapes within and across the covariate space.

Let ${\boldsymbol x}$ be a vector of random covariates and $T$ the positive-valued
survival response variable. 
We model the joint response-covariate density using a DP mixture model:
\begin{eqnarray}\label{eqn:dpmm}
f(t, {\boldsymbol x} \mid G) = \int k(t, {\boldsymbol x} \mid {\boldsymbol \theta}) \,
\text{d}G( {\boldsymbol \theta});  \ \ \  G \sim \text{DP}(\alpha, G_0) 
\end{eqnarray}
where $k(t, {\boldsymbol x} \mid {\boldsymbol \theta})$ is the joint
kernel density for survival time and covariates, 
and the mixing distribution, $G$, is assigned a DP prior \citep{ferguson}. 
The model is completed with hyperpriors for the DP precision parameter,
$\alpha$, 
and for (some of) the parameters
of the baseline (centering) distribution $G_{0}$.
Under the DP constructive definition \citep{sethuraman}, a realization $G$ 
from DP$(\alpha, G_0)$ is almost surely of the form $\sum_{l=1}^\infty w_l\delta_{\boldsymbol{\theta}_l}$, 
where the atoms are independently and identically distributed (i.i.d.)
from the baseline distribution, $\boldsymbol{\theta}_l
\stackrel{\text{i.i.d.}}{\sim} G_{0}$, with the weights constructed through 
stick-breaking: $w_1= v_1$ and $w_l=$ $v_l\prod_{r=1}^{l-1} (1-v_r)$, for $l \geq 2$, where 
$v_l \stackrel{\text{i.i.d.}}{\sim} \text{Beta}(1,\alpha)$ (independently of the $\boldsymbol{\theta}_l$).

Hence, the density in (\ref{eqn:dpmm}) can be re-written as $f(t, {\boldsymbol x} \mid G) =$ 
$\sum_{l = 1}^\infty w_l k(t, {\boldsymbol x} \mid {\boldsymbol \theta}_l)$. Directly from their 
definitions, the conditional response density can be expressed as 
$f(t \mid {\boldsymbol x}, G)=$
$\sum_{l=1}^{\infty}  q_l({\boldsymbol x}; {\boldsymbol \theta}_l) \, 
k(t \mid {\boldsymbol x}, {\boldsymbol \theta}_l)$, and the conditional survival function as 
\begin{equation}\label{DPM_survival}
S(t \mid {\boldsymbol x}, G) = \sum_{l=1}^{\infty}  q_l({\boldsymbol x}; {\boldsymbol \theta}_l) \, 
S(t \mid {\boldsymbol x}, {\boldsymbol \theta}_l)
\end{equation}
where $q_l({\boldsymbol x}; {\boldsymbol \theta}_l) = w_l k({\boldsymbol x} \mid {\boldsymbol \theta}_l) / \{\sum_{r=1}^\infty w_r k({\boldsymbol x} \mid {\boldsymbol \theta}_r)\}$. Therefore, the conditional 
density and survival functionals are represented as mixtures of the corresponding kernel functions with 
covariate-dependent mixture weights. Analogously, the mean regression function is
$\text{E}(T \mid {\boldsymbol x}, G) =$ $\sum_{l=1}^\infty q_l({\boldsymbol x}; {\boldsymbol \theta}_l) 
\text{E}(T \mid {\boldsymbol x},{\boldsymbol \theta}_l)$ (a sufficient condition for finiteness of the
conditional expectation is provided later). The covariate-dependent mixture weights allow for 
local adjustment over the covariate space, thus enabling general shapes for the conditional 
response distribution and for the mean regression functional.

Importantly for our objective, this local mixture structure extends to the MRL functional. 
Using the form for the conditional survival function in (\ref{DPM_survival}) and the 
definition of the MRL regression function from (\ref{eqn:mrl}), we obtain
\begin{eqnarray}\label{eqn:dpmm_mrl}
m(t \mid {\boldsymbol x}, G) = \frac{\int_t^\infty S(u \mid {\boldsymbol x},G) \, 
\text{d}u}{S(t \mid {\boldsymbol x},G)} = 
\sum_{l=1}^{\infty} q^{*}_{l}(t,{\boldsymbol x} ;{\boldsymbol \theta}_l) \, 
m(t \mid {\boldsymbol x},{\boldsymbol \theta}_l)
\end{eqnarray}
where $q^{*}_{l}(t, {\boldsymbol x} ; {\boldsymbol \theta}_l) = w_l k({\boldsymbol x} \mid {\boldsymbol \theta}_l) 
S(t \mid {\boldsymbol x}, {\boldsymbol \theta}_l)/\{\sum_{r=1}^{\infty} w_{r} k({\boldsymbol x} \mid 
{\boldsymbol \theta}_r) S(t \mid {\boldsymbol x}, {\boldsymbol \theta}_r)\}$, and 
$m(t \mid {\boldsymbol x},{\boldsymbol \theta})$ is the MRL function of the conditional 
response distribution under the kernel. (Implicit here is the assumption that, under the kernel 
distribution, $\text{E}(T \mid {\boldsymbol x},{\boldsymbol \theta}) < \infty$, for any ${\boldsymbol x}$).
Therefore, our prior model for the MRL regression function admits a representation as a 
weighted sum of the conditional MRL functions associated with the kernel components, 
with weights that are dependent on both time and the covariate values. Important to 
note in the form of the mixture weights is that there are separate functions controlling 
the local adjustment over covariate values and time.    
Aside from the useful interpretation, expression (\ref{eqn:dpmm_mrl}) suggests the capacity 
of the model to capture non-standard MRL regression relationships over time, as well as  
general MRL function shapes across the covariate space.

We next turn to the choice of the DP mixture kernel, $k(t, {\boldsymbol x} \mid {\boldsymbol \theta})$. 
A structured approach to specifying dependent kernel densities involves a marginal density for the 
covariates, $k({\boldsymbol x} \mid {\boldsymbol \theta}_{1})$, and a parametric regression model 
for $k(t \mid {\boldsymbol x},{\boldsymbol \theta}_{2})$, where ${\boldsymbol \theta}=$
$({\boldsymbol \theta}_{1},{\boldsymbol \theta}_{2})$. For our data illustrations, we use the 
simpler form for the kernel density that corresponds to independent components for the survival 
response and the covariates, $k(t, {\boldsymbol x} \mid {\boldsymbol \theta})=$
$k({\boldsymbol x} \mid {\boldsymbol \theta}_{1}) k(t \mid {\boldsymbol \theta}_{2})$.
In this case, the prior model in (\ref{eqn:dpmm_mrl}) becomes a mixture of the marginal kernel
MRL functions, with weights that are still dependent on both time and covariate values through 
distinct functions, here, $S(t \mid {\boldsymbol \theta}_l)$ and 
$k({\boldsymbol x} \mid {\boldsymbol \theta}_l)$, respectively. As can be seen from the 
model structure and also demonstrated with the data examples, such a kernel density form 
strikes a good balance between inference flexibility and model complexity with respect to the 
dimensionality of the mixing parameter vector ${\boldsymbol \theta}$.
Regarding $k({\boldsymbol x} \mid {\boldsymbol \theta}_{1})$, when all the covariates are continuous, 
the multivariate normal density is a convenient choice, possibly after transformation for the values of 
some of the covariates. A normal kernel density can also accommodate ordinal categorical covariates 
through latent continuous variables \citep[e.g.,][]{deyoreokottas}. Alternatively, categorical covariates 
(whether ordinal or nominal) can be incorporated by adding a corresponding component to the kernel 
in a product form, or if relevant, through marginal and conditional densities for the continuous and 
categorical covariates \citep[e.g.,][]{taddy:kottas}.

A key consideration for the specification of $k(t \mid {\boldsymbol \theta}_{2})$ is to ensure that 
the MRL function $m(t \mid {\boldsymbol x}, G)$ is well defined, that is, we require that 
$\text{E}(T \mid {\boldsymbol x},G)$ is (almost surely) finite, for any ${\boldsymbol x}$. The 
following lemma (whose proof is given in Appendix A) provides a sufficient condition for 
finiteness of this conditional expectation.

\vspace{0.11cm}
\noindent 
{\bf Lemma.} Consider the DP mixture model in (\ref{eqn:dpmm}) with kernel of the general form 
$k({\boldsymbol x} \mid {\boldsymbol \theta}_{1}) k(t \mid {\boldsymbol x},{\boldsymbol \theta}_{2})$,
and with DP centering distribution $G_{0}({\boldsymbol \theta}_{1},{\boldsymbol \theta}_{2})=$
$G_{10}({\boldsymbol \theta}_{1}) G_{20}({\boldsymbol \theta}_{2})$,
and let ${\boldsymbol x}$ be a generic set of covariate values. 
If $\text{E}(T \mid {\boldsymbol x},{\boldsymbol \theta}_{2}) =$
$\int_{\mathbb{R}^{+}} u \, k(u \mid {\boldsymbol x},{\boldsymbol \theta}_{2}) \, \text{d}u 
< \infty$, and $\int \text{E}(T \mid {\boldsymbol x},{\boldsymbol \theta}_{2}) \,
\text{d}G_{20}({\boldsymbol \theta}_{2}) < \infty$, then, 
$\text{E}(T \mid {\boldsymbol x},G) < \infty$, almost surely.

\vspace{0.11cm}

Note that defining $G_{0}$ through independent components for ${\boldsymbol \theta}_{1}$ 
and ${\boldsymbol \theta}_{2}$ is a natural modeling strategy. Also, under independent kernel 
components for the response and covariates, 
$\text{E}(T \mid {\boldsymbol x},{\boldsymbol \theta}_{2})$ simplifies to the expectation 
of the kernel for survival time, and the second lemma condition becomes 
$\int \text{E}(T \mid {\boldsymbol \theta}_{2}) \, \text{d}G_{20}({\boldsymbol \theta}_{2}) < \infty$. 
For this model version, it is straightforward to verify the lemma conditions for the gamma density 
under a particular selection for $G_{20}$.
The gamma choice is unique in this respect among standard lifetime distributions in that 
it suffices for existence of the mixture MRL function without the need for awkward restrictions 
on the parameter space for ${\boldsymbol \theta}_{2}$. Further support for the gamma kernel choice
is provided by the fact that it generates both increasing and decreasing MRL functions (for shape 
parameter $< 1$ and $> 1$, respectively), its MRL function can be
expressed in a form that is easy to compute (see Section \ref{DPM_MRL_inference}), 
as well as by a denseness result for MRL functions corresponding to
gamma mixture distributions, obtained under the setting without covariates \citep{poynor:kottas}. 
We use the following parameterization for the gamma density, $k(t \mid {\boldsymbol \theta}_{2}) \equiv$ 
$k(t \mid \eta,\phi) \propto$ $t^{e^\eta -1}\text{exp}(-e^\phi t)$, with 
$(\eta, \phi)\in \mathbb{R}^2$, to facilitate selection of a dependent $G_{20}(\eta,\phi)$ 
distribution, taken to be bivariate Gaussian. Finally, we note that the lemma conditions remain 
generally easy to verify if one wishes to extend the gamma kernel density to depend on covariates, 
for instance, such that its mean is extended to $\exp(\eta - {\boldsymbol x}^T{\boldsymbol \beta})$.

\subsection{Posterior inference}
\label{DPM_MRL_inference}

We obtain samples from the posterior distribution of the DP mixture
model using the blocked Gibbs sampler \citep{ishwaran}.
In particular, the Markov chain Monte Carlo (MCMC) posterior simulation
method builds from a truncation approximation 
to the mixing distribution, $G_{L}=$ $\sum_{l=1}^{L} p_{l} \delta_{\boldsymbol{\theta}_l}$, with 
$\boldsymbol{\theta}_l \stackrel{\text{i.i.d.}}{\sim} G_{0}$, for $l=1,...,L$, 
$p_{l}=w_{l}$, for $l=1,...,L-1$, and $p_L=$ $1- \sum_{l=1}^{L-1} p_{l}$.  
The truncation level $L$ can be chosen to any desired level of accuracy, 
using DP properties. For instance, the prior expectation for the partial sum of the 
original DP weights, $\text{E}(\sum_{l=1}^{L} p_{l} \mid \alpha)=$
$1 - \{ \alpha/(\alpha + 1) \}^{L}$, can be averaged over the
hyperprior for $\alpha$ to estimate $\text{E}(\sum_{l=1}^{L} p_{l})$
for any value of the truncation level. Appendix B includes details of
the MCMC algorithm for the DDP mixture model developed in Section \ref{sec:ddp},
an algorithm that includes as a special case the one for the DP
mixture model used for the simulation examples of Section \ref{sec:dpmmsim}.

%
%
  
Posterior inference for the density, survival, and mean regression 
functionals can be obtained by evaluating the corresponding
conditional response distribution functional under model
(\ref{eqn:dpmm}) at any time $t$ and covariate values ${\boldsymbol x}$ 
of interest. The expressions for $f(t \mid {\boldsymbol x}, G)$,
$S(t \mid {\boldsymbol x}, G)$, and $\text{E}(T \mid {\boldsymbol x}, G)$
are computed using the posterior 
samples for $G_{L}$, thus involving finite sums at the inference stage. 
Posterior samples for the MRL regression function can be efficiently
computed using expression (\ref{eqn:dpmm_mrl}), provided the kernel
MRL function can be readily computed. This is indeed the case for the
gamma kernel distribution whose MRL function can be expressed in terms
of the Gamma function, $\Gamma(a)$, and the gamma distribution 
survival function, $S_{\Gamma}(t)$ \citep{govilt}. More specifically,
under the gamma density parameterization given in Section \ref{DPM_for_MRL},
$$
m(t \mid \eta,\phi) = 
\frac{ t^{e^{\eta}} \exp(- e^{\phi} t) \exp\{ \phi (e^{\eta} - 1) \} }
{ \Gamma(e^{\eta}) S_{\Gamma}(t \mid \eta,\phi) } + \exp(\eta - \phi) - t.
$$
This expression suffices for the model built from independent kernel
components for the survival response and covariates, and it can be easily
extended to accommodate a gamma kernel density that depends on covariates.

%
%
%
%

\subsection{Simulation examples}
\label{sec:dpmmsim}

We provide two simulation examples to demonstrate the model's 
capacity to capture a variety of MRL functional shapes. 
Both examples involve a single continuous covariate.
For the first example, we work with a finite mixture for the joint 
response-covariate distribution, specified such that the MRL function 
takes on various non-standard shapes at different parts of the covariate space. 
In the second example, we consider an exponentiated Weibull distribution
\citep{mudholkar} for the survival responses. This is a three-parameter 
extension of the Weibull distribution that achieves more general
shapes for the hazard rate and MRL function. The regression model for the 
simulation truth is built by defining the three response distribution parameters 
through specific functions of covariate values, which are drawn
from a uniform distribution. 
The two simulation scenarios are designed to correspond to a setting 
similar to the model structure, as well as a much more structured
parametric setting for the data generating stochastic mechanism.
We work with relatively large sample sizes ($1500$ and $500$ for the
first and second example) so that the data
sets provide reasonably accurate representations of the simulation truth, thus
rendering comparison with true MRL functions meaningful. The synthetic
data examples of Section \ref{simulation_DDPM} and the analysis of 
the real data in Section \ref{sec:lung_data} illustrate model
inferences under smaller sample sizes.

We apply the same DP mixture model to both synthetic data sets, with
mixture kernel defined through the product of the gamma 
density for the survival response, 
$k(t \mid \eta,\phi) \propto$ $t^{e^\eta -1}\text{exp}(-e^\phi t)$, 
and a normal density for the covariate, 
$\text{N}(x \mid \beta,\kappa^2) \propto$ $\text{exp}(-0.5\kappa^{-2}(x - \beta)^2)$.
The DP centering distribution is defined by 
$G_{0}(\eta,\phi,\beta,\kappa^2)=$
$\text{N}_{2}((\eta,\phi) \mid {\boldsymbol \mu}, {\boldsymbol \Sigma}) \,
\text{N}(\beta \mid\lambda, \tau^2) \, \Gamma^{-1}(\kappa^2 \mid a, \rho)$, 
where $\Gamma^{-1}(c,d)$ denotes the inverse-gamma distribution with
mean $d/(c-1)$ (provided $c>1$).
The model is completed with the following hyperpriors:
${\boldsymbol \mu} \sim \text{N}_2(a_\mu, B_\mu)$, 
${\boldsymbol \Sigma} \sim \text{IWish}(a_\Sigma, B_\Sigma)$,
$\lambda \sim  \text{N}(a_{\lambda}, b_{\lambda})$, 
$\tau^2 \sim \Gamma^{-1}(a_\tau, b_\tau)$,
$\rho \sim \Gamma(a_\rho, b_\rho)$, and 
$\alpha \sim \Gamma(\alpha \mid a_\alpha, b_\alpha)$, 
where $\Gamma(c,d)$ denotes the gamma distribution with mean $c/d$.
For both examples, we set $a_\alpha = 3$, $b_\alpha = 0.1$, and 
$L=80$ for the DP truncation level.

\subsubsection{Simulation 1}

\begin{figure}[t!]
 \centering
          \includegraphics[height=1.7in,width=2.2in]{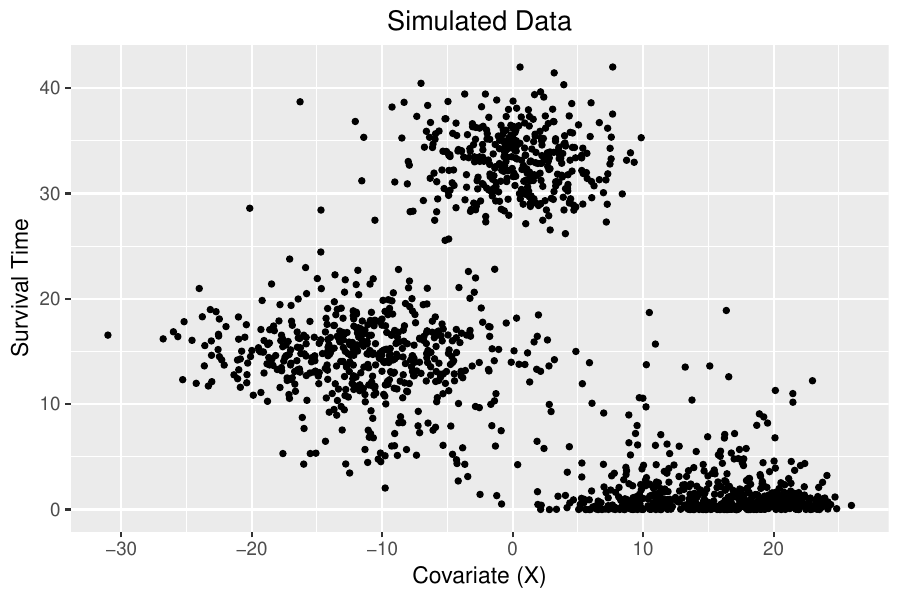}
          \includegraphics[height=1.7in,width=2.2in]{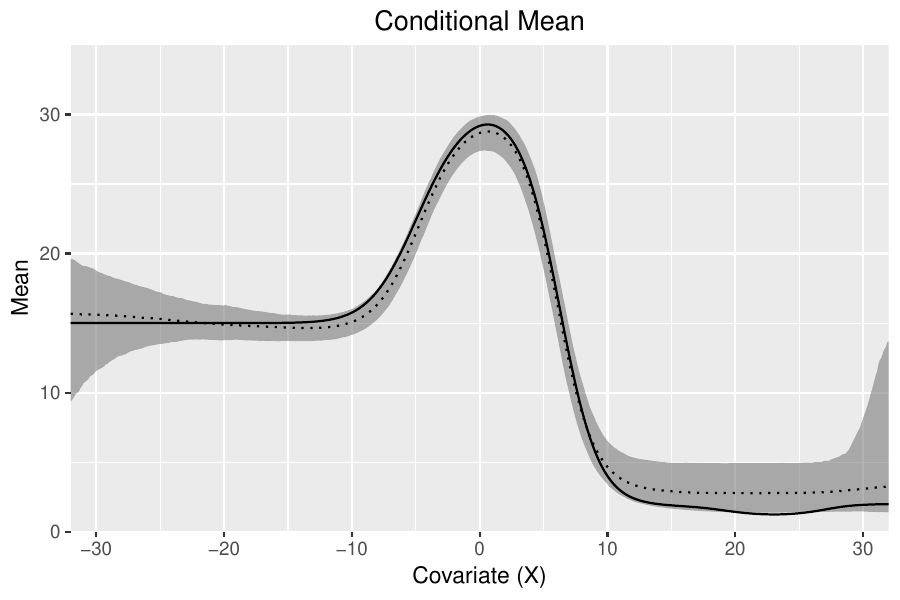}
\caption{Simulated data from the finite mixture. The left panel plots
the data. The right panel shows point (dotted line) and interval
estimates (gray bands) of $\text{E}(T \mid x,G)$, overlaid on the true 
conditional expectation (solid line).}
\label{fig:curvefitsim}
\end{figure}

We simulate $1500$ observations from a population with density:
$f(t,x)=$ $\sum_{l=1}^{6} q_l\Gamma(t\mid a_l, b_l) \text{N}(x \mid m_l,s^2_l)$,
where $\{ a_{l} \}=$ $(45, 3, 125, 0.4,0.5, 4)$, $\{ b_{l} \}=$
$(3,0.2, 3.8, 0.2,0.3,5)$, $\{ m_{l} \}=$
$(-12, -8, 0 ,12, 18, 21)$, $\{ s_{l} \}=$ $(6, 5, 4, 5, 3, 2)$, and
$\{ q_{l} \}=$ $(0.28, 0.1, 0.25, 0.21, 0.11, 0.05)$. The simulated data is shown
in the left panel of Figure \ref{fig:curvefitsim}.  The following
hyper priors were assumed: $a_\mu= (0.59, -2.12)$, $B_\mu=B_\Sigma =
((0.019,0)', (0, 0.019)')$, $a_\lambda=0$, $a_\tau = 2$, $a_\rho=1$,
$b_\lambda=b_\tau=88$, $b_\rho=1/88$.

\begin{figure}[t!]
 \centering
           \includegraphics[height=1.7in,width=2in]{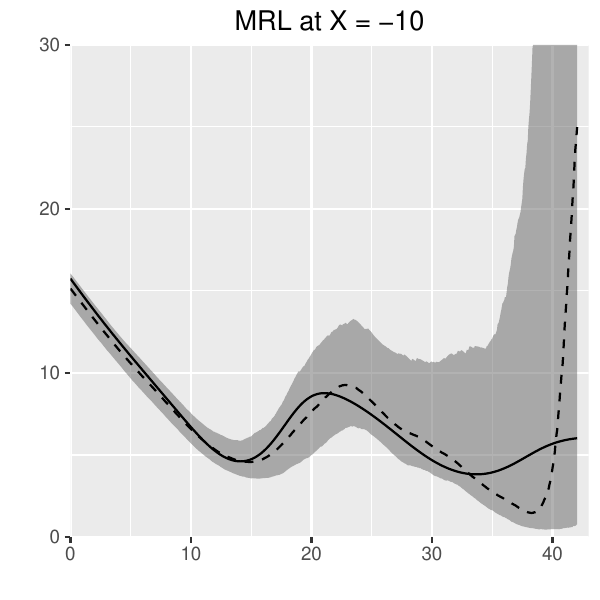} 
           \includegraphics[height=1.7in,width=2in]{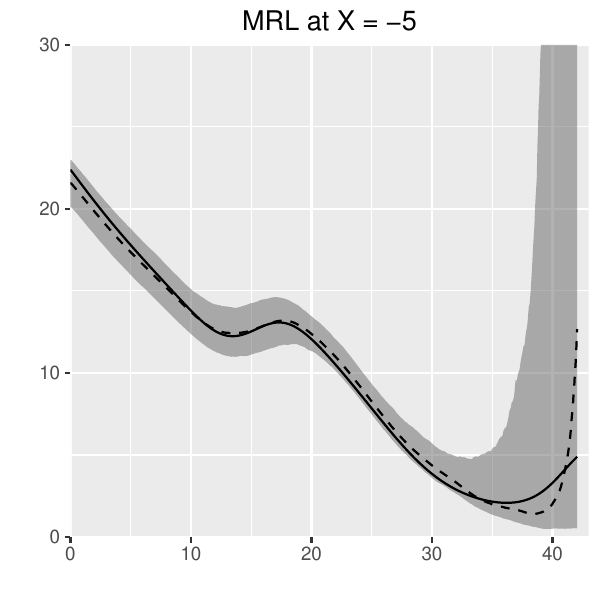}\\
            \includegraphics[height=1.7in,width=2in]{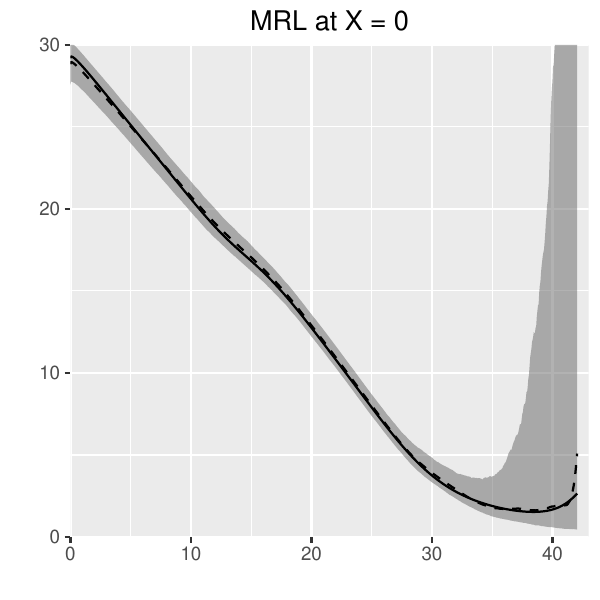}
           \includegraphics[height=1.7in,width=2in]{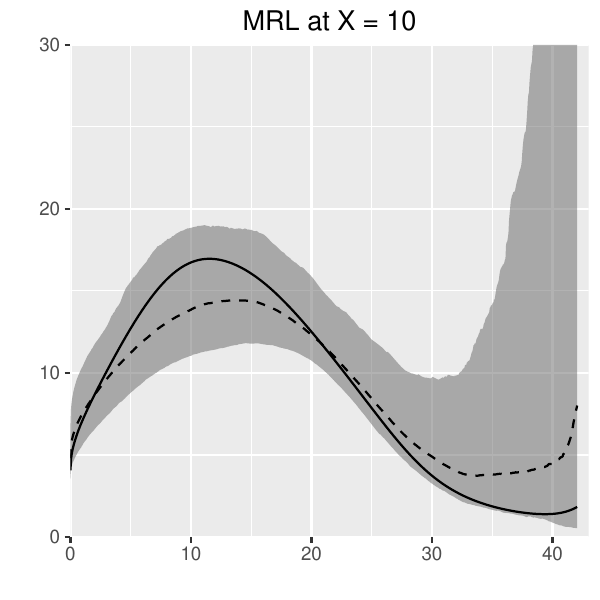}\\ 
           \caption{Simulated data from the finite mixture.
Point (dashed line) and $95\%$ interval estimates (gray bands) of the MRL 
function for the specified covariate value overlaying the true MRL
function of the population (solid line).}\label{fig:curvefitmrl}
\end{figure}

The mean of the survival times across a grid of covariate values is shown in Figure~\ref{fig:curvefitsim} (right panel).  
In general, the model is able to capture the non-linear trend of the mean over the covariate values.  The truth is captured within the $95\%$ interval estimate save for a small sliver barely outside the interval near the right tail of the covariate space where data is sparse. 
The results for MRL functional inference is shown in Figure~\ref{fig:curvefitmrl}.  We provide point and $95\%$ interval estimates for the MRL function at four different covariate values.  The model is able to capture the overall shape of the true MRL functions, despite the variety of and often complexity of the shapes.  At covariate values where the data is most dense, such as $x=-5$ and $x=0$, the inference is more precise as is seen in the narrow interval bands.  As we move to covariate values where data is more sparse, the wide interval bands reflect the uncertainty of the MRL functional shape.

\subsubsection{Simulation 2}

\begin{figure}[t!]
 \centering
          \includegraphics[height=1.7in,width=2.2in]{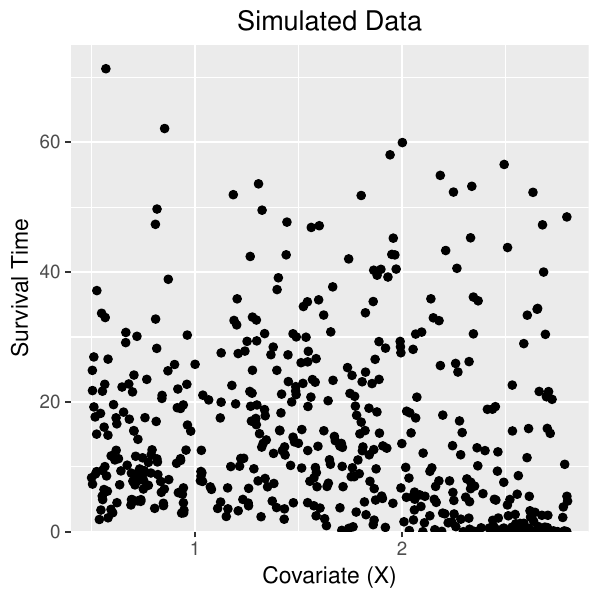}
          \includegraphics[height=1.7in,width=2.2in]{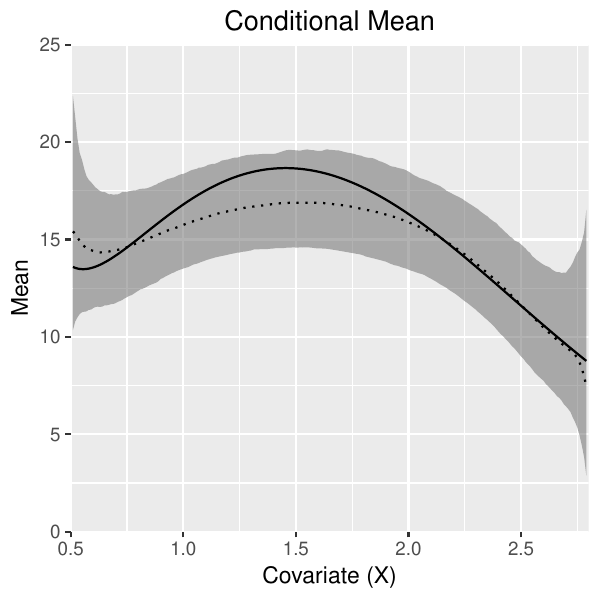}
\caption{Simulated data from the exponentiated Weibull regression model. 
The left panel plots the data. The right panel shows point (dotted line) and interval
estimates (gray bands) of $\text{E}(T \mid x,G)$, overlaid on the true 
conditional expectation (solid line).}
\label{fig:ExpWeibsim}
\end{figure}

The exponentiated Weibull population has
survival function, $S(t \mid \alpha', \theta', \sigma' )=$
$1- [1- \text{exp}\{ - (t/\sigma')^{\alpha'}\}]^{\theta'}$.  The MRL function
associated with this distribution can take on increasing, decreasing,
constant, upside-down bathtub, and bathtub shapes depending on the
shape parameters, $\alpha'$ and $\theta'$, as well as their product
($\sigma'$ is a scale parameter).  We sample $500$ observations from
an Exponentiated Weibull population with $\alpha' = X$, $\theta' =
\text{exp}(2.93 -1.96X)$, and $\sigma' = 14\text{log}(X^3 + 1)$, where
$X \sim \text{Unif}(0.5, 2.8)$.  The simulated data is shown in the
left panel of Figure \ref{fig:ExpWeibsim}.  The following hyper priors
were assumed: $a_\mu= (2.0, -0.8)$, $B_\mu=B_\sigma = ((0.11,0)', (0,
0.11)')$, $a_\lambda=0$, $a_\tau = 2$, $a_\rho=1$,
$b_\lambda=b_\tau=4.6$, $b_\rho=1/4.6$.

\begin{figure}[t!]
 \centering
           \includegraphics[height=1.7in,width=2in]{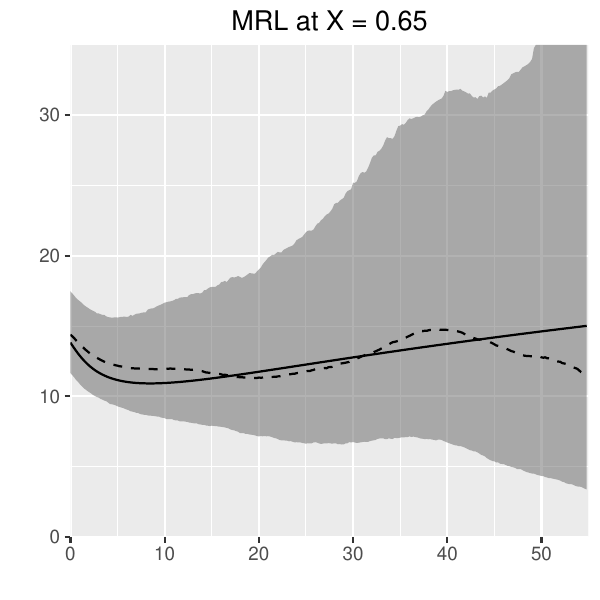} 
           \includegraphics[height=1.7in,width=2in]{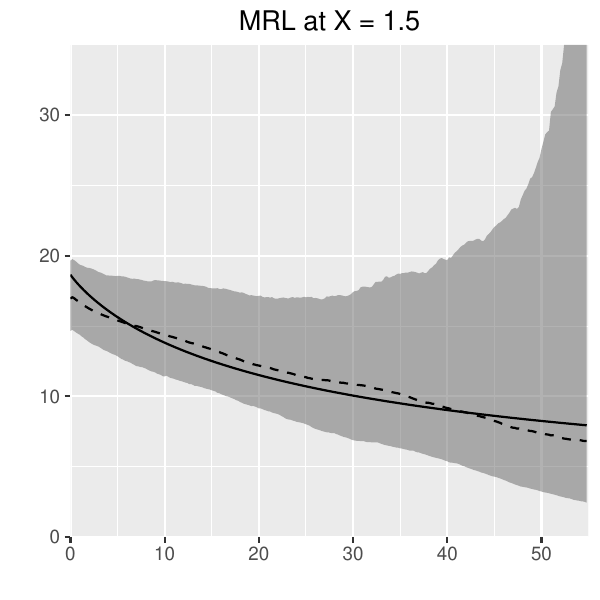}\\
            \includegraphics[height=1.7in,width=2in]{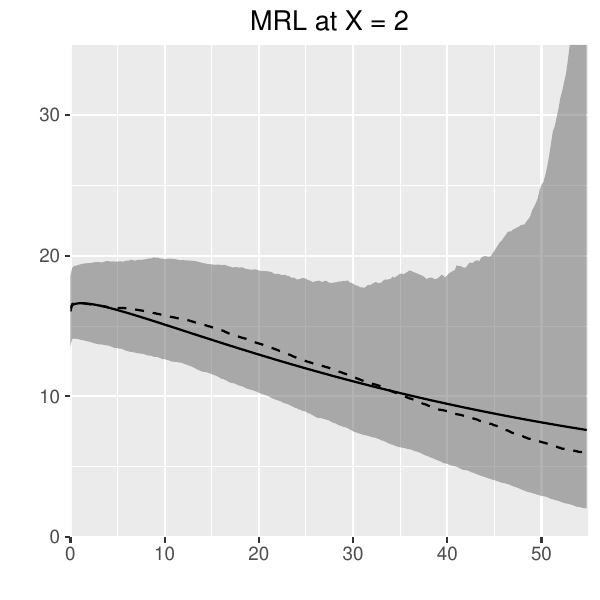}
           \includegraphics[height=1.7in,width=2in]{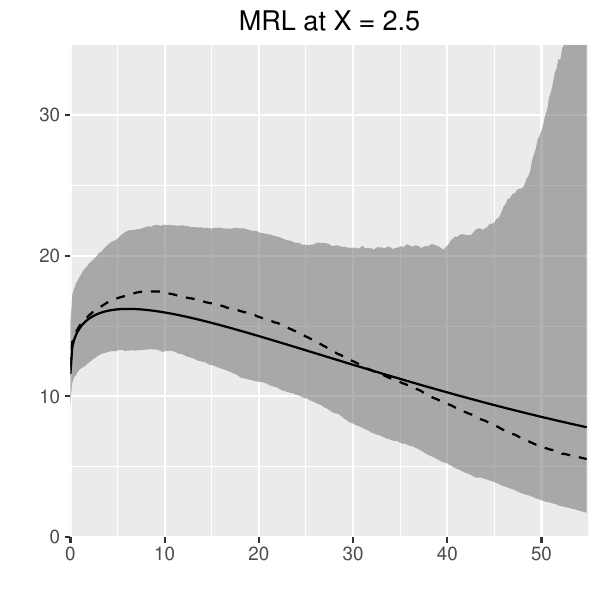}\\ 
\caption{Simulated data from the exponentiated Weibull regression model.
Point (dotted line) and $95\%$ interval estimates (gray bands) of the MRL
function for the specified covariate value overlaying the true MRL
function of the population (solid line).}\label{fig:ExpWeibmrl}
\end{figure}

The mean of the survival times across a grid of covariate values is
shown in Figure~\ref{fig:ExpWeibsim} (right panel).  Once again, the
true mean regression exhibits a non-linear trend that is increasing
until about $x = 1.5$ then decreases.  The is captured well within the
$95\%$ interval estimate and the parabolic shape is clearly mimicked
by the point estimate. The results for MRL functional inference is
shown in Figure~\ref{fig:ExpWeibmrl} at four covariate values.  In all
four scenarios, the truth is captured within the $95\%$ interval
bands while the general shapes are mimicked by the point estimates.

%
%

\section{Dependent DP mixture model for MRL regression}\label{sec:ddp}

\subsection{The DDP mixture model formulation}

Often in clinical trials, researchers are interested in modeling
survival times of patients under treatment and control groups. Since
the underlying population pre treatment is typically the same,
it is reasonable to expect that the survival distributions of the two
groups exhibit similarities.  Thus modeling groups jointly is a
natural choice, offering potential learning for the correlation as
well as borrowing inferential strength across groups.  We propose to
do so by generalizing the DP mixture model described in Section
\ref{sec:curve_reg} to a dependent DP (DDP) mixture model.  Under this
framework, we achieve non-standard shapes in the MRL regression
functions, that may even differ across groups contingent on the
strength of the dependence across experimental groups.

Let $s\in S$ represent in general the index of dependence.  In our
case, this indicates the experimental group, that is $S=\{T,C\}$ where
($T$) is the treatment group and ($C$) is the control group.  The
model in (\ref{eqn:dpmm}) can be extended to $f(t, {\boldsymbol x}\mid
G_{s}) = \int_{\boldsymbol \Theta} k(t,{\boldsymbol x}\mid
{\boldsymbol \theta})dG_s({\boldsymbol \theta})$ for  $s \in S$, where
now we are modeling a pair of dependent random mixing distributions
$\{G_s : s \in S\}$.  We seek to model the distributions in such a
way as to incorporate dependencies across experimental groups, while
maintaining marginally the DP prior, $G_s \sim \text{DP}$, for each
$s\in S$. \citet{mac} develops the dependent DP prior in generality
with both the weights and atoms under the stick-breaking definition
dependent on experimental group: $G_s =$
$ \sum_{l=1}^\infty \omega_{ls}\delta_{{\boldsymbol \theta}_{ls}}$.  Marginally,
$G_s \sim \text{DP}(\alpha_s, G_{0s})$ for each $s \in S$.
\citet{mac} goes on to describe the computational difficulties in
modeling dependencies in the weights across groups, thus motivating
development of the common weights model.  In this model, the weights
do not change over the groups, only the locations vary, $G_s=$
$\sum_{l=1}^\infty \omega_{l}\delta_{{\boldsymbol \theta}_{ls}}$.  
Applications of common weights DDP models include \cite{deiorio04},
\cite{rodriguez:horst}, \cite{deiorio09}, \citet{kottas:behseta}, 
and \cite{kottas:fronczyk}.

While computationally convenient and a useful extension of the basic DP prior, assuming the same weights 
has potential disadvantages in our setting.  A practical disadvantage of the common weights DDP construction 
involves applications with a moderate to large number of covariates.  
For such cases, the common weights prior requires 
building dependence across $s\in S$ for a large number of kernel parameters, whereas modeling dependence 
through the weights is not affected by the dimensionality of the mixture kernel.  In situations where we might 
expect similar locations across groups, modeling dependence through the weights is more attractive.  
In our context, we may expect the two groups to be comprised of similar components which however exhibit
different prevalence across survival time.

We thus use mixing distributions of the form $G_s =$ $\sum_{l=1}^\infty w_{ls}\delta_{{\boldsymbol \theta}_l}$,
for $s \in \{ T,C\}$ representing the treatment and control groups,
respectively, and the DDP mixture model becomes
\begin{eqnarray}
f(t, {\boldsymbol x}\mid G_{s}) &=& \int_{\boldsymbol \Theta} k(t,{\boldsymbol x}\mid {\boldsymbol \theta})dG_s({\boldsymbol \theta}); \ \ \ G_s \sim \text{DDP}({\boldsymbol \Phi}, G_0)   \label{eqn:ddpmm}
\end{eqnarray}
where ${\boldsymbol \Phi}$ represents the parameters associated with
the construction of the dependent weights of $G_s$. The  common atoms
are defined, as usual, arising i.i.d. from the baseline distribution, $G_0$.  
%
%
It follows that the conditional response density can be written as $f(t \mid {\boldsymbol x},  G_s) =$
$ \sum_{l=1}^\infty w_{ls} k(t\mid {\boldsymbol x},{\boldsymbol \theta}_l)$, and the conditional survival function as
\begin{eqnarray}\label{eqn:ddpsurv}
S(t \mid {\boldsymbol x}, G_s) &=& \sum_{l=1}^\infty q_{ls}({\boldsymbol x}; {\boldsymbol \theta}_l)S(t \mid {\boldsymbol x}, {\boldsymbol \theta}_l)
\end{eqnarray}      
where $q_{ls}({\boldsymbol x}; {\boldsymbol \theta}_l) =
w_{ls}k({\boldsymbol x} \mid {\boldsymbol
  \theta}_l)/\{\sum_{r=1}^\infty  w_{rs}k({\boldsymbol x} \mid
{\boldsymbol \theta}_r)\}$.  Likewise, the mean regression function is
$ E(t\mid{\boldsymbol x}, G_s)  = \sum_{l=1}^\infty
q_{ls}({\boldsymbol x}; {\boldsymbol \theta}_l) E(t\mid{\boldsymbol
  x},{\boldsymbol \theta}_l)$.  Thus, the conditional density,
conditional survival, and mean regression functions are weighted
mixtures of the corresponding kernel functions with weights dependent
on the covariate as well as the group. 
This structure implies that general shapes are tractable not 
only across the covariate space, but also across the groups. 

Using the conditional survival form of (\ref{eqn:ddpsurv}) under 
definition (\ref{eqn:mrl}), the MRL regression function is written as 
\begin{eqnarray}
m(t\mid {\boldsymbol x}, G_s) = \frac{\int_t^\infty S(u\mid {\boldsymbol x},G_s) \text{d}u}{S(t\mid {\boldsymbol x},G_s)} 
= \sum_{l=1}^\infty q^*_{ls}(t,{\boldsymbol x} ;{\boldsymbol \theta}_l)m(t\mid {\boldsymbol x},{\boldsymbol \theta}_l)\label{eqn:ddpmm_mrl}
\end{eqnarray}
\noindent where $q^*_{ls}(t, {\boldsymbol x} ; {\boldsymbol \theta}_l)
= w_{ls} k({\boldsymbol x}\mid {\boldsymbol \theta}_l)S(t\mid
{\boldsymbol x}, {\boldsymbol \theta}_l)/\{\sum_{l=1}^L w_{ls}
k({\boldsymbol x}\mid {\boldsymbol \theta}_l)S(t\mid{\boldsymbol x},
{\boldsymbol \theta}_l)\}$.  Here, we see that the local weighted
mixture structure is again extended to the MRL regression, and the
local adjustments over the covariates, time, and (now) groups each
have separate controlling terms in the mixture weights.  We have
already demonstrated the flexibility of the MRL regression function
within and across the covariate space under form (\ref{eqn:dpmm_mrl}).
We preserve that same flexibility under the form in
(\ref{eqn:ddpmm_mrl}) for a specific group $s$ with the addition of
the model's ability extract information across groups while
maintaining the unique features within groups.  Indeed, the MRL
regression function can vary in shape across 
the groups at the same covariate value if the data suggests.

Next, we turn to the construction of the dependent weights of
$G_s$. Under the stick-breaking method in obtaining the weights, we
sample independently the latent parameters, $\upsilon_l \sim
\text{Beta}(1,\alpha)$, which is equivalent to using $\zeta_l =(1-
\upsilon_l) \sim \text{Beta}(\alpha, 1)$   for  $l\in\{1, 2,...\}$.
If we use a bivariate beta distribution for $(\zeta_{Tl},\zeta_{Cl})$
having $\text{Beta}(\alpha, 1)$ marginals, we can incorporate the
dependence between the two groups while maintaining the DP 
prior marginally for each group.
Specifically, the weights are defined as follows: $w_{1s} = 1 -
\zeta_{1s}, \ w_{ls} = (1- \zeta_{ls})\prod_{r=1}^{l-1} \zeta_{rs}$
for  $l\in\{2,3,...\}$, with $(\zeta_{lC}, \zeta_{lT})\mid
{\boldsymbol \Phi} \stackrel{\text{ind}}{\sim} \mbox{Biv-Beta}(\cdot
\mid {\boldsymbol \Phi})$, a bivariate beta distribution such that
marginally the $\zeta_{lC}$ and $\zeta_{lT}$ are
$\text{Beta}(\alpha,1)$ distributed.

%
%


We work with a bivariate beta distribution from \citet{kotz}, defined 
constructively through products of independent beta distributed random
variables. In particular, to define the bivariate beta distribution
for $(X,Y)$, start with independent random variables,
$U\sim \text{Beta}(a_1,b_1)$, $V\sim \text{Beta}(a_2,b_2)$, and $W\sim
\text{Beta}(b,c)$, subject to the constraint, $c=a_1 +b_1 = a_2+b_2$.
Then, define $X = UW$ and $Y = VW$.  The
marginals are given by $X \sim \text{Beta}(a_1, b_1 + b)$ and $Y\sim
\text{Beta}(a_2, b_2 +b)$.  We can obtain the desired beta marginals
for  $\zeta_{Cr}$ and $\zeta_{Tr}$ by setting $b_1+b=$ $b_2+b=1$.
We also take $a_{1}=a_{2}$ such that the random mixing
distributions have the same marginal DP prior. The joint
density of $(X,Y)$ has a complicated form, but it can be sampled from using
latent variables.  The correlation has an analytic expression, and it
can be shown to be positive. 
Induced correlations in the model under this bivariate beta
distribution are discussed in Section~\ref{sec:prop_ddpmm} below.

\subsection{Properties of the DDP mixture model}\label{sec:prop_ddpmm}

In this section, we study the correlation structure induced by the
bivariate beta distribution given in the previous section.  Under this bivariate beta
construction, the correlation is driven by both parameters, $\alpha$ and $b$.
The construction is based off of the product of independent beta
distributions.  Recall, we start with sampling the independent latent
variables: $U\sim\text{Beta}(\alpha, 1-b)$, $V\sim\text{Beta}(\alpha,
1-b)$, $W\sim\text{Beta}(\alpha +1 -b, b)$.  Let $\zeta_C = UW$ and
$\zeta_T = VW$.  The weights are defined by $w_{s1}=1-\zeta_{1s}$,
$w_{ls}=(1-\zeta_{ls})\prod_{r=1}^{l-1}\zeta_{rs}$, for
$l\in\{2,3,...\}$.  The correlation structures for the latent
variables a well as the weights are detailed in the Appendix.

We are interested in obtaining the correlation between the two mixing distributions, $G_C$ and $G_T$, implied under this bivariate beta distribution.  Let $B \in {\boldsymbol \Theta}$ represent a subset of the space of the mixing parameters.  In the model we present, ${\boldsymbol \Theta}$ is equivalent to $\mathbb{R}^2$, so $B$ is simply a subset of $\mathbb{R}^2$.  Recall that the mixing distribution for group $s$ has form $G_s(B)= \sum_{l=1}^\infty w_{ls}\delta_{{\boldsymbol \theta}_l}(B)$.  Marginally, $G_s(B)$ follows a DP, so the expectation and variance of $G_s(B)$ is $G_0(B)$ and $G_0(B)[1-G_0(B)]/(\alpha+1)$, respectively.  The covariance between $G_C(B)$ and $G_T(B)$ is given by $Cov\left(\sum_{l=1}^\infty w_{lC} \delta_{{\boldsymbol \theta}_l}(B),\sum_{l=1}^\infty w_{lT} \delta_{{\boldsymbol \theta}_l}(B)\right)$, which boils down to the expression, $G_0(B)\sum_{l=1}^\infty w_{lC}w_{lT} + 2G_0^2(B)\sum_{l=1}^\infty \sum_{m=l+1}^\infty w_{lC}w_{mT} -G_0^2(B)$.  The infinite series converges under geometric series, and the covariance simplifies to be:
{\small \begin{eqnarray*}
\text{Cov}(G_C(B),G_T(B)) &=& G_0(B)(1-G_0(B))\left(\frac{(\alpha-2)b+\alpha +2}{\alpha(2\alpha -3b+5) -2b+2} \right)\end{eqnarray*}}

\noindent The correlation, therefore, does not depend on the choice of $B$ or $G_0$, it is driven by $\alpha$ and $b$ alone:
{\small \begin{eqnarray}
\text{Corr}(G_C(B),G_T(B)) &=& \frac{(\alpha+1)((\alpha-2)b+\alpha +2)}{\alpha(2\alpha -3b+5) -2b+2} \label{eqn:cor}
\end{eqnarray}}

The correlation of the mixing distribution lives on the interval
$(1/2, 1)$.  As $\alpha \to 0$ and/or $b\to 1$, the correlation tends
to $1$.  When $\alpha \to \infty$ the correlation tends to $(b+1)/2$
and as $b \to 0$ the correlation tends to $(\alpha+1)/(2\alpha+1)$, so
when $\alpha \to \infty$ and  $b \to 0$ the correlation goes to $1/2$.
Although this correlation space is limited, it is a typical range seen
in the literature (e.g. \cite{mckenzie}).  It can easily be shown that
the correlation of the survival distributions between the two groups
given $G_C$ and $G_T$ also live on $(1/2, 1)$, which demonstrates the
importance of prior knowledge of the relationship between the
distributions of the two group survival times.  While the possible
values of correlation on the distributions of the survival times are
restricted to $(1/2,1)$, the correlation between the survival times
across the two groups, $\text{Corr}(T_C, T_T)$, 
takes on values in $(0,1)$; see Appendix A for details.

\subsection{Synthetic data examples}
\label{simulation_DDPM}

\begin{figure}[t!]
 \centering
          \includegraphics[height=2in,width=2in]{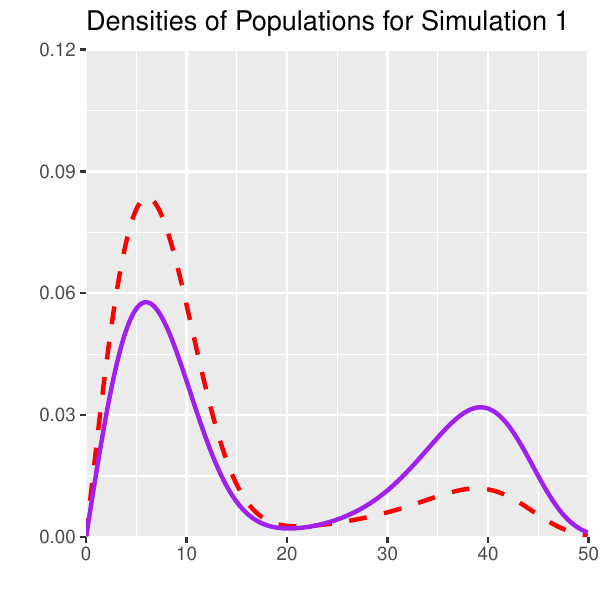}\includegraphics[height=2in,width=2in]{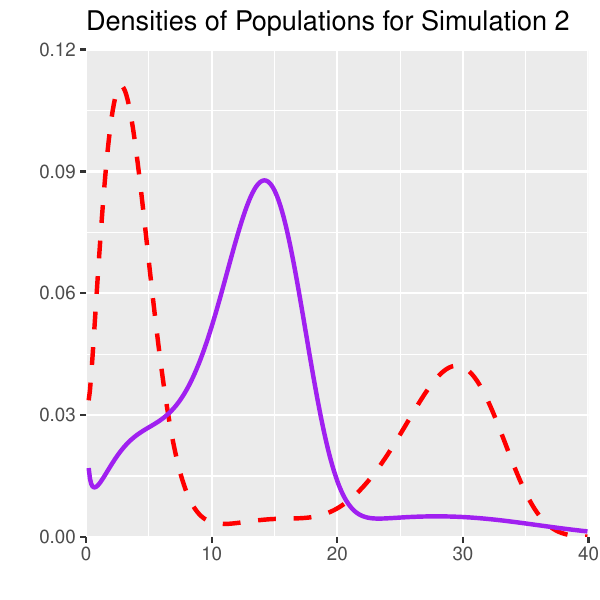}
        	 \caption{Simulation 1 population densities (left) and Simulation 2 population densities (right).  The red dashed curve represents the first population ($T_1$) while the purple solid represents the second ($T_2$) in each simulation.}\label{fig:DDPMMsimpop}
\end{figure}

In this section, we construct two sets of populations to investigate
the performance of the  DDP mixture model without covariates.  The
first set of populations is constructed using a mixture of Weibull
distributions having the same atoms and different weights.  We would
expect the DDP mixture model perform well under this scenario since
the population shares the same structure as the DDP in the model.  The
populations for the first simulation is shown in the left panel in
Figure \ref{fig:DDPMMsimpop}.  The panel shows how the two populations
look similar having modes at the same locations just differing
prevalences.  The second set of populations is also constructed using
a mixture of Weibull distributions, however, this time we use both
different weights and atoms.   The intention is to test the model's
inferential ability for populations that have quite different
features.  Figure \ref{fig:DDPMMsimpop} shows the density populations
of the second simulation in the right panel.  The second population
exhibits a single mode in between the two modes of the first 
population. The panel indicates that the two densities are quite dissimilar.


We assume the same distributional specifications in the DDP mixture
model for both simulations.  Namely, assume a gamma  kernel density
for $k(t\mid {\boldsymbol \theta}) = \Gamma(t \mid \eta, \phi)$ with
baseline distribution $G_{0}(\eta, \phi)=  \text{N}_2((\eta, \phi)\mid
{\boldsymbol \mu}, {\boldsymbol \Sigma})$. 
We specify the following priors: 
${\boldsymbol \mu} \sim \text{N}_2(\boldsymbol{\mu}\mid a_\mu, B_\mu),
{\boldsymbol \Sigma} \sim  \text{IWish}({\boldsymbol \Sigma}\mid
a_\Sigma, B_\Sigma), \alpha \sim \Gamma(\alpha \mid a_\alpha,
b_\alpha), \ b \sim \text{Unif}(b\mid 0,1)$. We obtain posterior
samples using the blocked Gibbs sampler \citep{ishwaran}
and working with the latent parameters of the bivariate beta
distribution. Posterior samples are based on a truncation
approximation, $G_{Ls}$, to $G_s$.  See Appendix B for details 
on the posterior sampling algorithm.

\subsubsection{Simulation 1}

In Simulation 1, we demonstrate the model's ability to perform under circumstances in which resembles the structure of our model. Specifically, we simulate from two Weibull mixture distributions that share mixture locations, but have different weights:  $T_1 \sim 0.7\text{Weib}(2,8) + 0.1\text{Weib}(3,10)  + 0.05\text{Weib}(4,30) + 0.15\text{Weib}(8,40) $ and $T_2 \sim 0.5\text{Weib}(2,8) + 0.05\text{Weib}(3,10)  + 0.025\text{Weib}(4,30) + 0.425\text{Weib}(8,40)$. The populations are comprised of four components each.  We sample $250$ survival times from the first population and $100$ survival times from the second population.  We do not consider censoring or covariates here. We place a uniform prior on the $b$ parameter and a gamma prior on $\alpha$ with shape parameter $ 2$ and rate parameter $0.8$.  The number of components is conservatively set at $40$.  We assume $a_\mu = (1.87, 0.25)'$, $B_\mu=b_\Sigma=((0.27, 0)',(0, 0.27)')$, and $a_\Sigma = 4$.  After burn in and thinning, we obtain 2000 independent posterior samples.

\begin{figure}[t!]
 \centering
          \includegraphics[height=1.75in,width=1.65in]{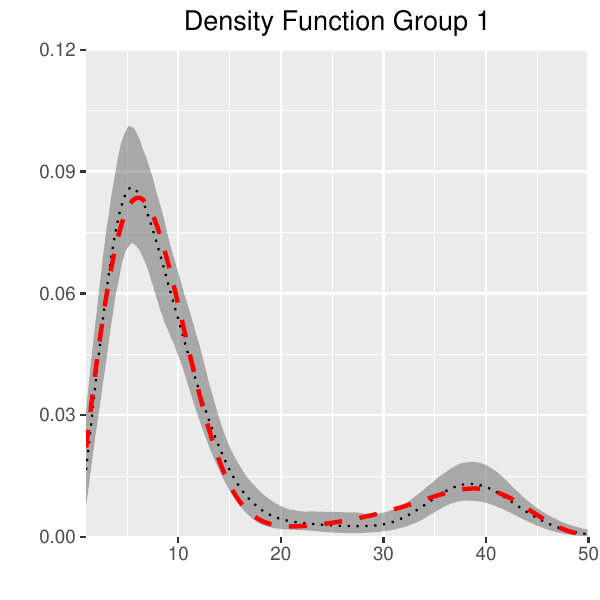}  \includegraphics[height=1.75in,width=1.65in]{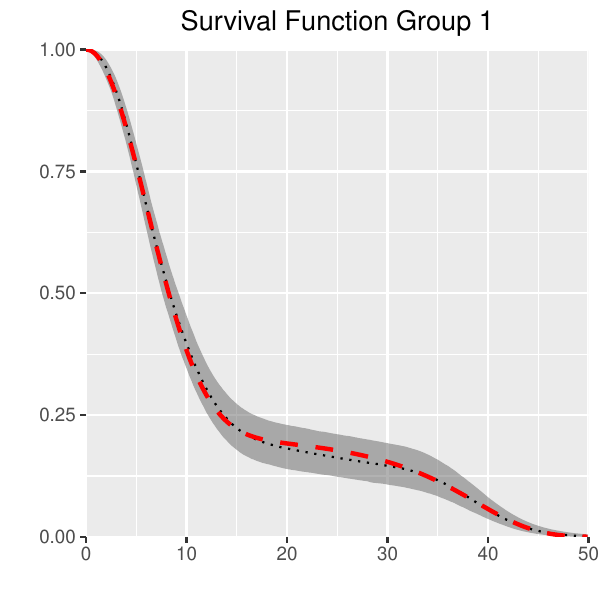} \includegraphics[height=1.75in,width=1.65in]{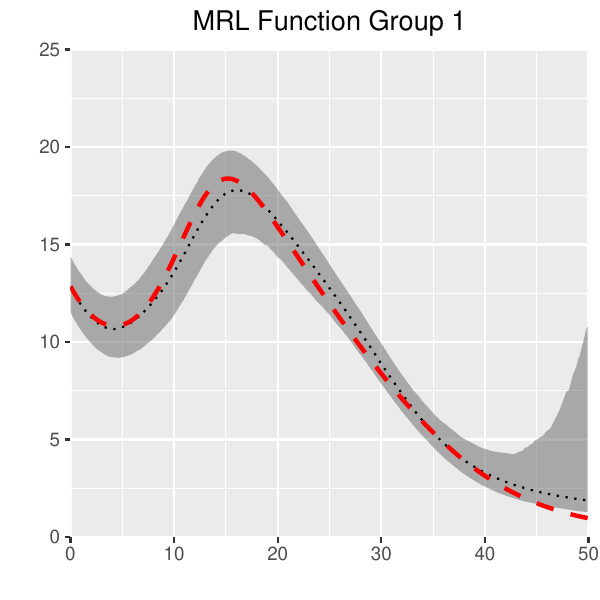} \\\includegraphics[height=1.75in,width=1.65in]{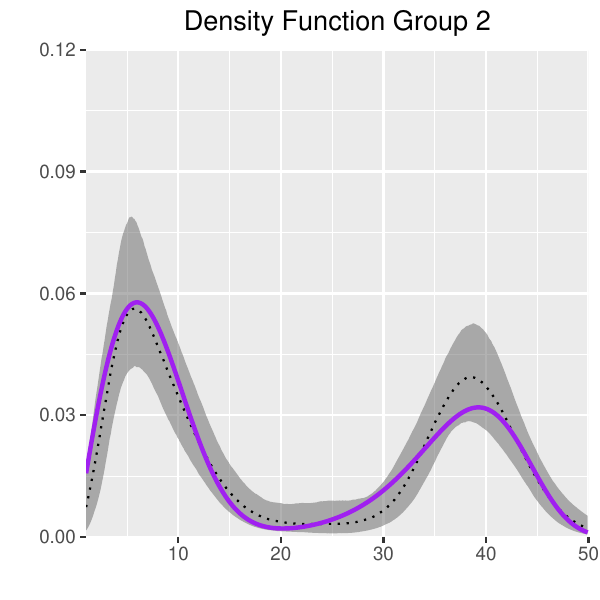}\includegraphics[height=1.75in,width=1.65in]{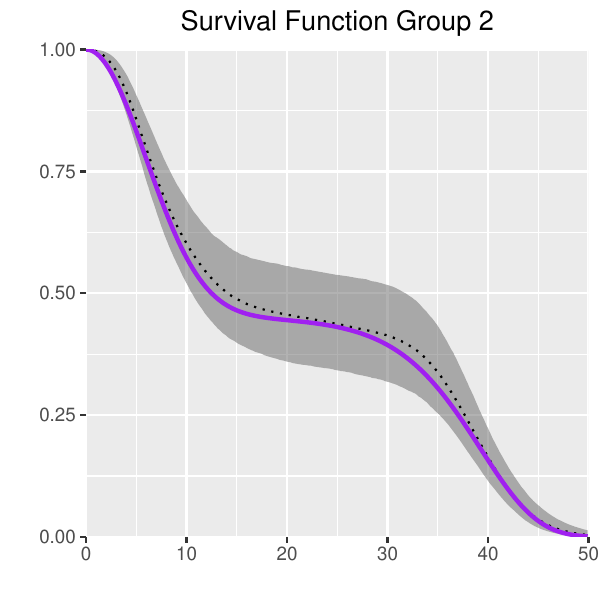} \includegraphics[height=1.75in,width=1.65in]{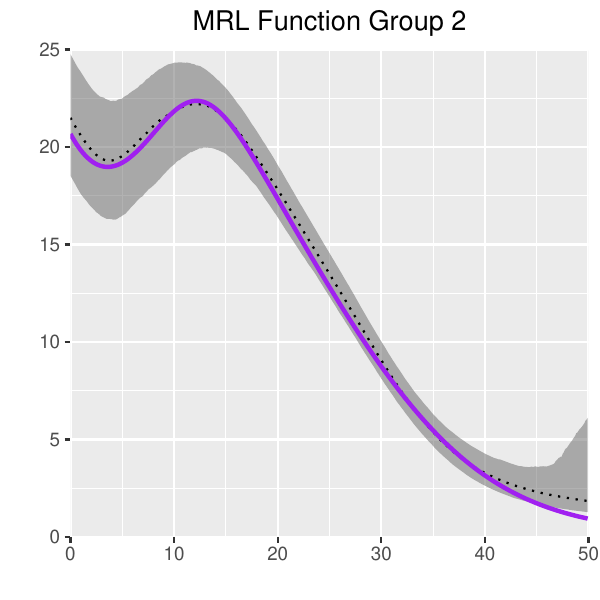}
        
        	 \caption{Simulation 1. Simulation 2. Posterior point and $95\%$ interval estimates for density (left), survival (middle), and MRL (right) functions. The truth is given by the dashed red (Group 1) and solid purple (Group 2).}\label{fig:DDPMMsim1fun}
\end{figure}

The $95\%$ posterior credible intervals for $\alpha$, $b$, and $\text{Corr}(G_C, G_T)$ are given by $(1.89, 14.45)$, $(0.15, 0.78)$, and $(0.59, 0.88)$, respectively.  Inference for the density, survival, and MRL functions are provided in Figure \ref{fig:DDPMMsim1fun}. The model is able to express the features of the functionals, and the true population density is captured within the $95\%$ interval estimates save for the very tail where data is very sparse.  In particular, the flexibility of the model is demonstrated in the MRL function.  The true MRL is non-standard in both groups: initially decreasing, followed by an increase after about time $5$, and then decreasing again after about time $12$.  The difference in sample size between the two groups is indicated by the slightly larger interval bands in Group 2 for the majority of the support of the data.

\subsubsection{Simulation 2}

The second simulation example is intended to be more of a challenge to the model.  The populations consist of mixtures of Weibull distributions, however, here we use different weights, locations, and number of components.  Group 1 is comprised of four components, while Group 2 is comprised of five:  $T_1 \sim 0.5\text{Weib}(2,4) + 0.05\text{Weib}(0.6,4)  + 0.025\text{Weib}(5,15) + 0.425\text{Weib}(8,30) $ and $T_2 \sim 0.02\text{Weib}(0.6,1) + 0.02\text{Weib}(2,4)  + 0.66\text{Weib}(5,15) + 0.2\text{Weib}(2,8)+ 0.1\text{Weib}(4,30)$. We simulate $250$ observations from each population.  All observations are fully observed, and no covariates are considered.  Once again, we use a uniform prior on $b$, and  gamma prior on $\alpha$ with shape parameter $2$ and rate parameter $0.8$.  The number of components is set at $40$, which is a conservative value for these data.  We assume $a_\mu =(3.02,  0.54)'$, $B_\mu=B_\Sigma=((0.1,0)',(0,0.1)')$, and $a_\Sigma = 4$.   After burn in and thinning, we obtain $2000$ independent posterior samples.  



\begin{figure}[t!]
 \centering
          \includegraphics[height=1.75in,width=1.65in]{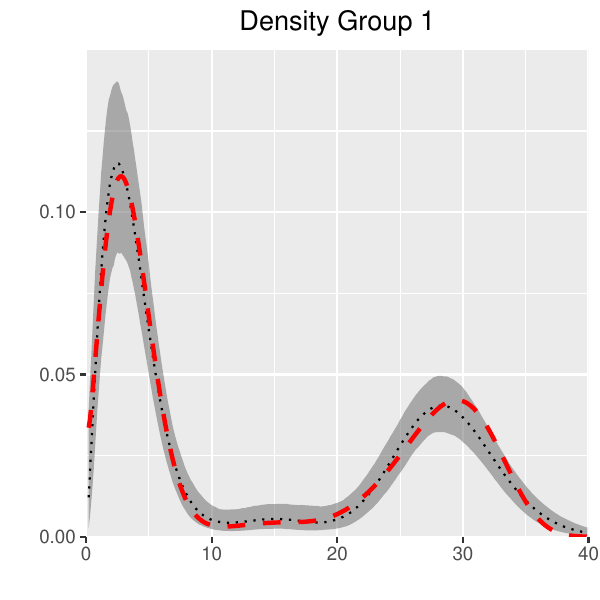}\includegraphics[height=1.75in,width=1.65in]{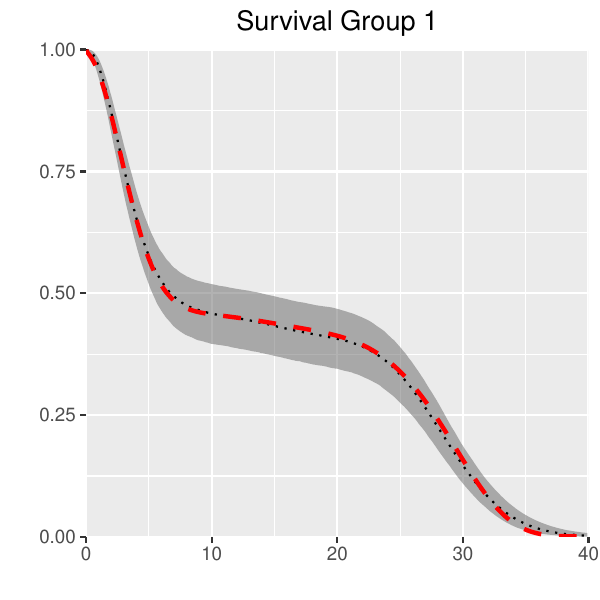}\includegraphics[height=1.75in,width=1.65in]{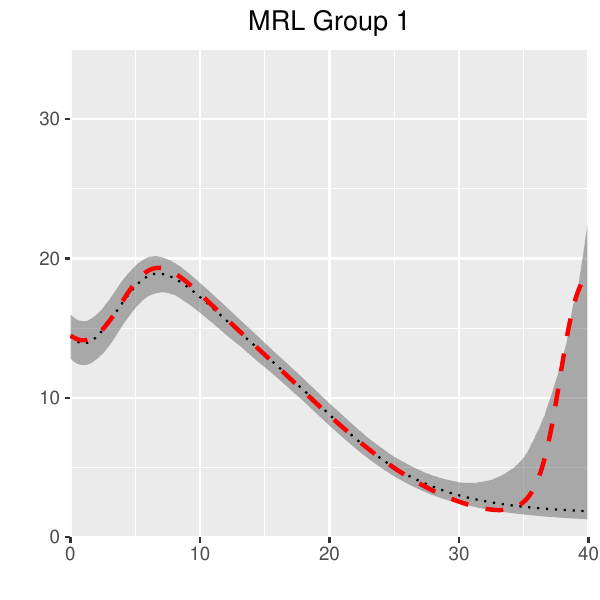}\\
          \includegraphics[height=1.75in,width=1.65in]{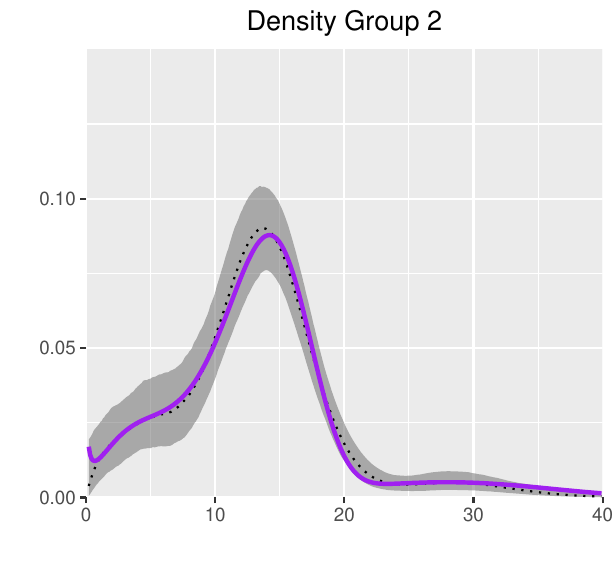}\includegraphics[height=1.75in,width=1.65in]{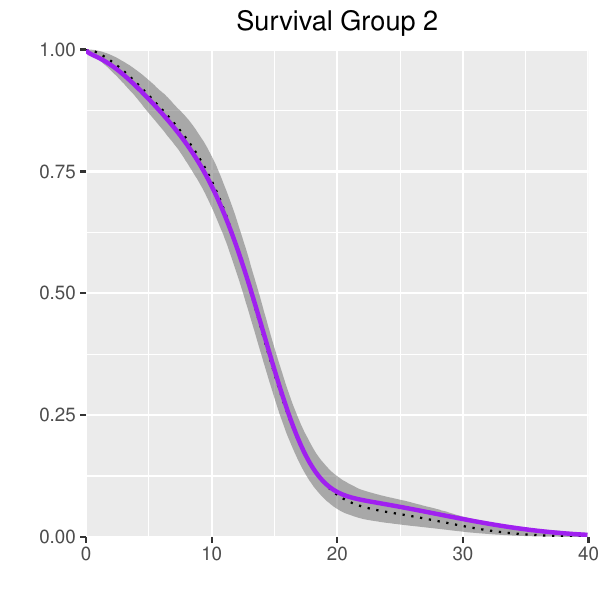}\includegraphics[height=1.75in,width=1.65in]{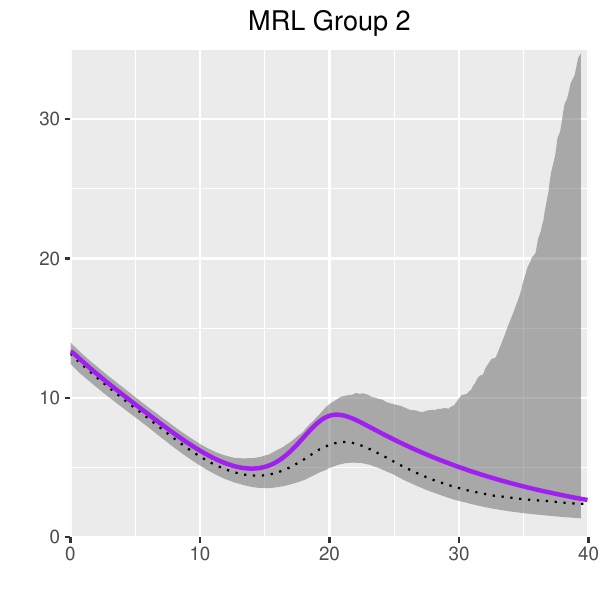}
        	 \caption{Simulation 2. Posterior point and $95\%$ interval estimates for density (left), survival (middle), and MRL (right) functions. The truth is given by the dashed red (Group 1) and solid purple (Group 2).}\label{fig:DDPMsim2post}
\end{figure}

The $95\%$ posterior credible intervals for $\alpha$, $b$, and $\text{Corr}(G_C, G_T)$ are given by $(0.76, 3.88)$, $(0.12, 0.72)$, and $(0.62, 0.84)$, respectively. The posterior results for the density, survival, and MRL functions are shown in Figure \ref{fig:DDPMsim2post}.  Despite the difference in the features of the functionals between the two groups, the model is able to capture the features of each group with accuracy.  This is especially exciting for the MRL functions.  The MRL functions are quite different from one another, and both are non-standard shapes.  The model has no problem capturing both shapes of the MRL functions. The only area where we can see struggle in the model for the MRL function inference is in the tails of the functionals.  The true MRL function of Group 1 is slightly above the upper interval estimate of the model.  This may be just due to the random nature of simulated data; this simulated data may suggest a lower MRL function in the tail.  Another possibility is the extreme difference between the MRL functions of the two groups in the tails.  Group 1 shoots up sharply, while Group 2 remains gradually decreasing.  A third contributor to the tail struggle is that the sparsity of the data in this area, so models in general a have a tougher time achieving accuracy.  Even with these elements against the model, the struggle is not significant.  
   
The results from the two simulations demonstrate the practical utility of the
DDP mixture model. The model is able to incorporate dependence across
two populations to achieve accurate inference in the functionals of
each population. In particular, the model provides
flexible MRL inference for two groups that exhibit MRL functions
with different features across the range of survival.

%
%

\section{Small cell lung cancer data example}\label{sec:lung_data}

We consider a dataset that comprises survival times, in days, of 
patients with small cell lung cancer under two treatment groups \citep{ying}. 
The patients were randomly assigned to one of two treatments referred to as Arm A and Arm B.  Arm A patients received cisplatin (P) followed by etoposide (E), while Arm B patients received (E) followed by (P).  Arm A consists of $62$ survival times, $15$ of which are right censored.  Arm B consists of $59$ survival times, $8$ of which are right censored.  The age of each patient upon entry is also available, however, in Section~\ref{sec:lung_nocov}, we will work with the treatment as the only covariate.  We later incorporate the age covariate in Section~\ref{sec:lung_cov}. 


\subsection{Comparison of experimental groups}\label{sec:lung_nocov}
\subsubsection{Results under DDP mixture model}

We fit a DDP mixture model with gamma kernel to these data.  Priors were specified using an analogous approach as described in \cite{poynor:kottas}, i.e., using the range and midrange of the observed survival times, which, in practice, would be specified by the expert.  We place a uniform prior on $b$ and a gamma prior with shape parameter $2$ and rate parameter $0.5$ is placed on $\alpha$, and set $L = 80$. The posterior $95\%$ credible intervals for $\alpha$ and $b$ are given by $(1.5, 11.9)$ and $(0.22,0.72)$, respectively.  We achieve some learning for $\alpha$ and a bit more for $b$.  Consequently, the model is able to learn about the correlation between the mixing distributions.  Using (\ref{eqn:cor}), we can obtain the posterior $95\%$ credible interval for $\text{Corr}(G_C,G_T)$ to be $(0.63, 0.85)$.  The posterior densities for both $\alpha$ and $b$ indicate learning for these parameters.   These data imply a fairly strong correlation between the mixing distributions as well as between the population distributions of the survival times under Arm A and Arm B.        

\begin{figure}[t!]
 \centering
          \includegraphics[height=2in,width=2.5in]{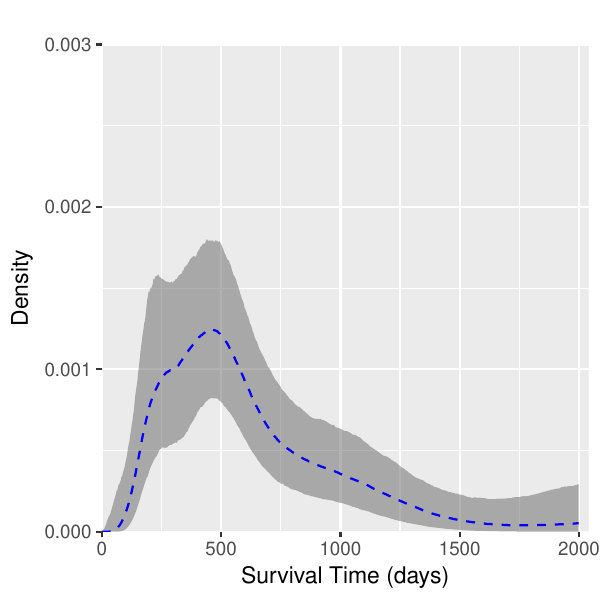}\includegraphics[height=2in,width=2.5in]{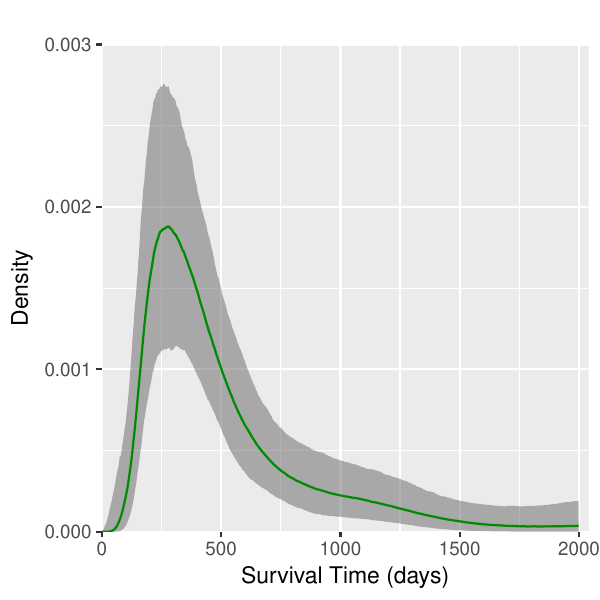}\\
          \includegraphics[height=2in,width=2.5in]{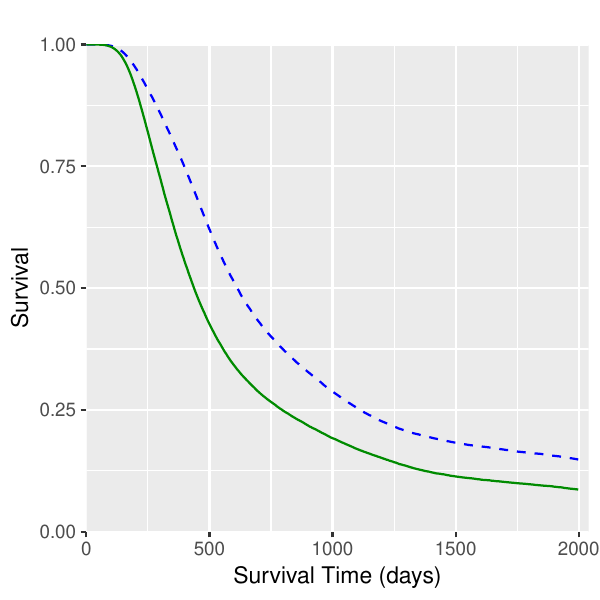}\includegraphics[height=2in,width=2.5in]{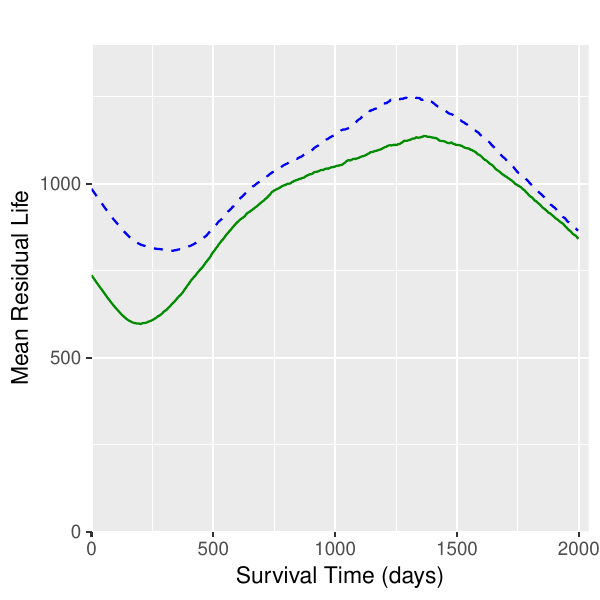}
        	 \caption{Small cell lung cancer data.
Posterior point and $95\%$ interval estimates of the density function
for Arm A (upper left) and Arm B (upper right).  Posterior point
estimate of the survival function (bottom left) and the mean residual
life function (bottom right) for Arm A (blue dashed) and Arm B (green
solid).}
\label{fig:DDPMMabMRL}
\end{figure}

Inference for the density, survival, and MRL functions are provided in Figure \ref{fig:DDPMMabMRL}.  The point estimates for the density have the same general shape to the point estimates obtained by \cite{kottas:krn}, who employ a semiparametric regression model.  Both models indicate a mode at about 450 days for Arm A and 350 days  for Arm B.  However, the point estimates under the DDP mixture model are smoother than under the semi-parametric regression model for both groups.  The difference is seen more obviously in the Arm B treatment.  The point estimates for the two survival curves indicates that Arm A has a higher survival rate across the range of the data starting from about 200 days.  The MRL regression exhibits a non-linear trend with Arm A having higher MRL over the entire time. When comparing the results under the DDP mixture model from under the independent DP mixture model, we see the same non-linear trend and favorability of Arm A over Arm B, however, the separation between the two groups is far less under DDP mixture model compared to the DP mixture model (see, Figure 3 in \cite{poynor:kottas}).  Arm B is the group that appears to be most affected  by the model change.  Specifically, the point estimate for Arm B is shifted up.  The shift is most drastic in the tail where data become more sparse.

\begin{figure}[t]
\centering
\includegraphics[height=2in,width=2.5in]{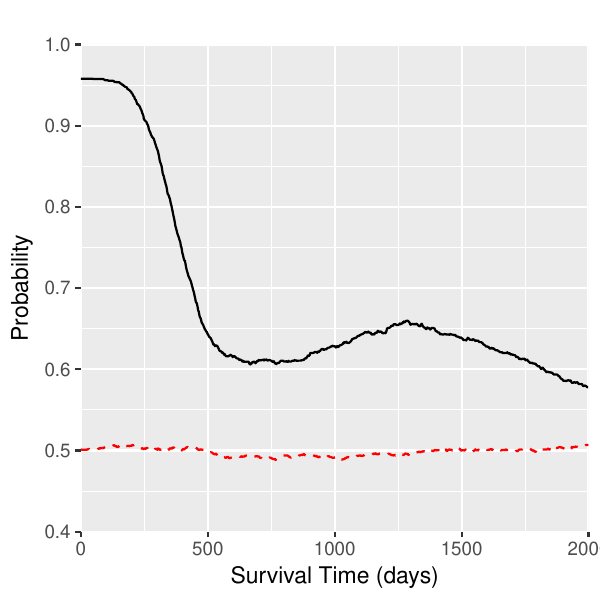}
\caption{Small cell lung cancer data. 
The posterior (black solid) and prior (red dashed)  probability 
of the MRL function of Arm A being higher than the MRL function of Arm B over a grid of survival times (days). 
}\label{fig:MRLdiffprobDDP}
\end{figure}

In Figure \ref{fig:MRLdiffprobDDP}, we look at the prior probability, 
$\text{Pr}(m_A(t) > m_B(t))$, and posterior probability, 
$\text{Pr}(m_A(t) > m_B(t) \mid \text{data})$, under the DDP mixture
model. This figure is analogous
to Figure 8 in \cite{poynor:kottas}. The prior probabilities under
both models do not favor one MRL function over the other at any time
point.  We also see from the figures that the posterior probability
changes in a similar fashion as we move across the time space.
Specifically, the probability is highest at smaller survival times
then dips down followed by an increase and then then tapers back down.
The range in probabilities is larger in Figure
\ref{fig:MRLdiffprobDDP}, with some probabilities reaching below
$0.6$.  In particular, Figure \ref{fig:MRLdiffprobDDP} indicates a
lower probability of the MRL function of Arm A being higher than the
MRL function of Arm B after about $500$ days.

\subsubsection{Model comparison}\label{sec:cpo}
 

In regards to model comparison, we are not aware of any competitive
models for inference on the MRL regression function.  However, the
small cell lung cancer data set has been used for illustration of
semiparametric survival regression models.  In particular,
\cite{kottas:krn} develop a Bayesian semiparametric model for
quantile regression, based on a linear quantile regression function
and a non-parametric scale mixture of uniform densities for the error
distribution.
Therefore, we formally compare the predictive performance of
the DDP mixture model for these 
data using the CPO criterion and comparing to the summary values 
reported in \cite{kottas:krn}.

The CPO of the $i^{th}$ observation $\text{CPO}_i$ can be expressed in
terms of the joint posterior distribution of the model
parameters,  ${\boldsymbol \Psi}$, given {\it all} the observations:
$\text{CPO}_i = \left(\int f(t_i|{\boldsymbol \Psi}, x_i)^{-1}
  \pi({\boldsymbol \Psi}| data)d{\boldsymbol \Psi}\right)^{-1}$. The
expression often does not have a closed form, so MCMC
approximation is used (see, for example, \cite{chen:shao}). The DDP
mixture model requires a slightly different expression for the CPO
values.  We provide the 
expression and derivation details in Appendix C.

A summary of the CPO values were obtained by averaging over the
log-CPO values, ALPML, in each group. The ALPML that are reported in
\cite{kottas:krn} include $-6.91$ for the non-parametric 
scale mixture of uniform densities, and $11.56$ for a
Weibull proportional hazards model. The ALPML of the DDP mixture model
is $-6.05$,  indicating better predictive performance compared to
these models.


\subsection{Incorporating the age covariate}\label{sec:lung_cov}

Here, we incorporate the age (in years) of the subjects, upon entrance
into the study, that is also available in the small cell lung cancer
dataset.  The researchers did not select subjects from particular
ages, so it is not a fixed covariate, and it can thus be incorporated 
into the model through a joint response-covariate distribution.

\begin{figure}[t!]
 \centering
          \includegraphics[height=2in,width=2.5in]{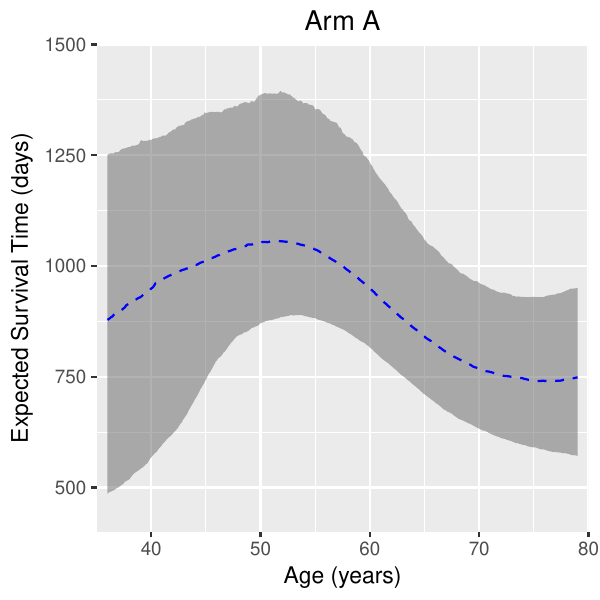}\includegraphics[height=2in,width=2.5in]{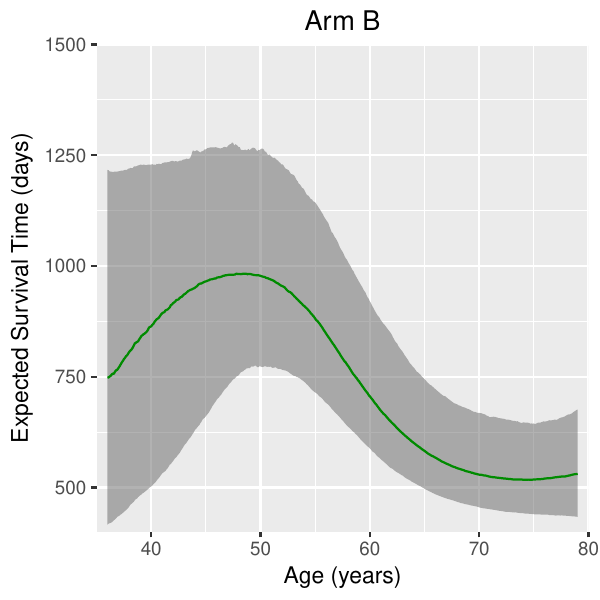}\\
        	 \caption{Small cell lung cancer data.
Point and $80\% $ interval estimates of the conditional mean of the
survival distribution of Arm A (blue dashed) and Arm B (green solid)
across a grid of age values (in years).}
\label{fig:condmean}
\end{figure}

In Figure \ref{fig:condmean}, we plot the mean regression function
over a grid of ages.  Recall that the mean regression is a weighted
sum of the kernel component means.  Moreover, the weights are
functions of the covariate, indicating the potential of the model to
capture non-standard relationships across the covariate space.  This
ability is demonstrated in Figure \ref{fig:condmean} where we see an
increase in the mean survival from about age $36$ to just after $50$,
followed by a steeper decline, 
particularly in Arm B, and then leveling out at higher ages. 

\begin{figure}[t!]
 \centering
          \includegraphics[height=1.8in,width=1.72in]{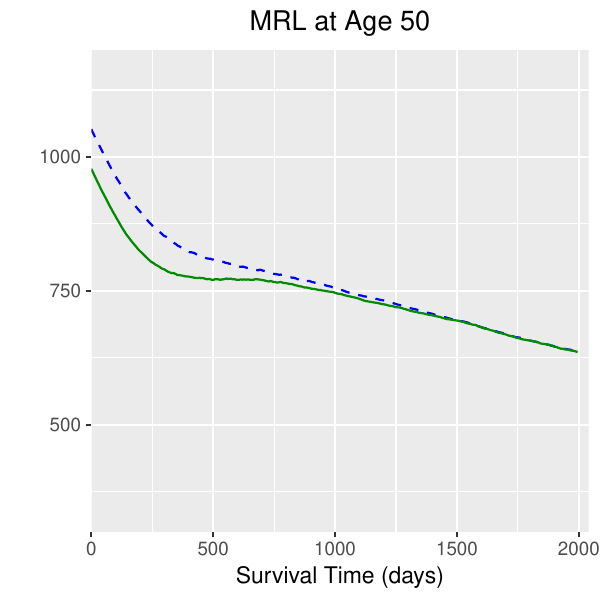}\includegraphics[height=1.8in,width=1.72in]{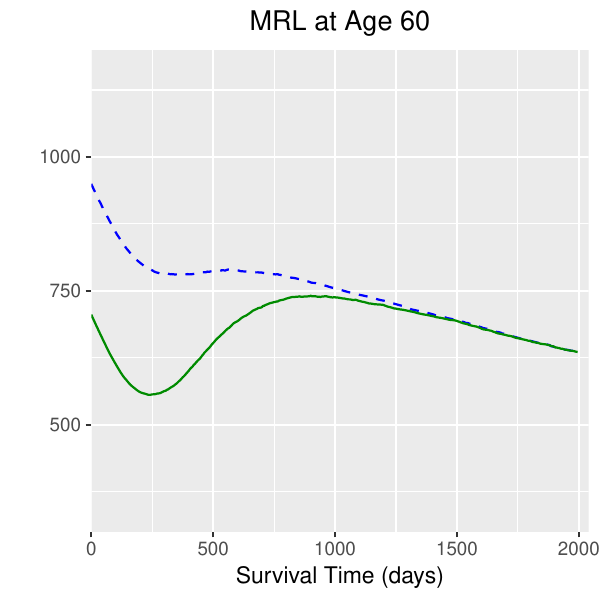}\includegraphics[height=1.8in,width=1.72in]{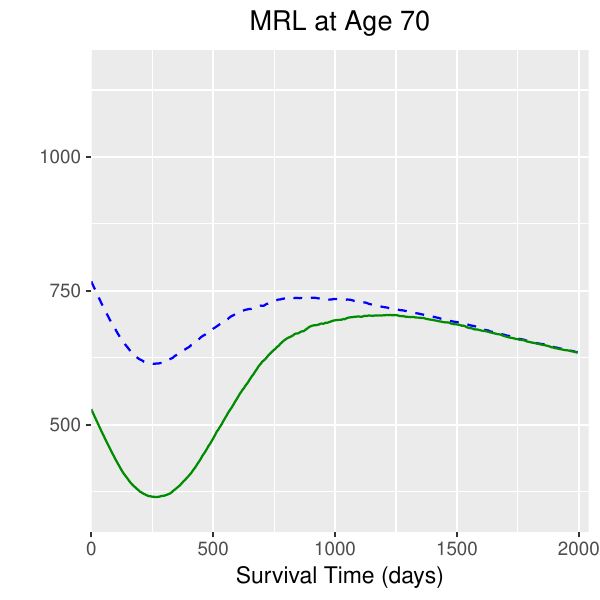}\\
           \includegraphics[height=1.8in,width=1.72in]{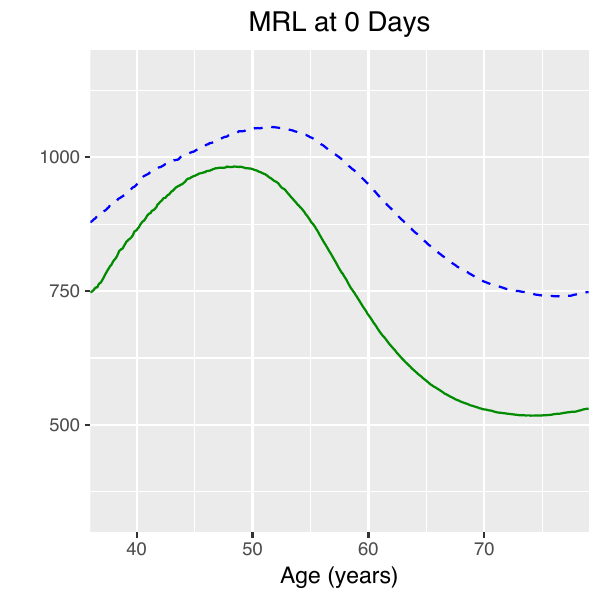}  \includegraphics[height=1.8in,width=1.72in]{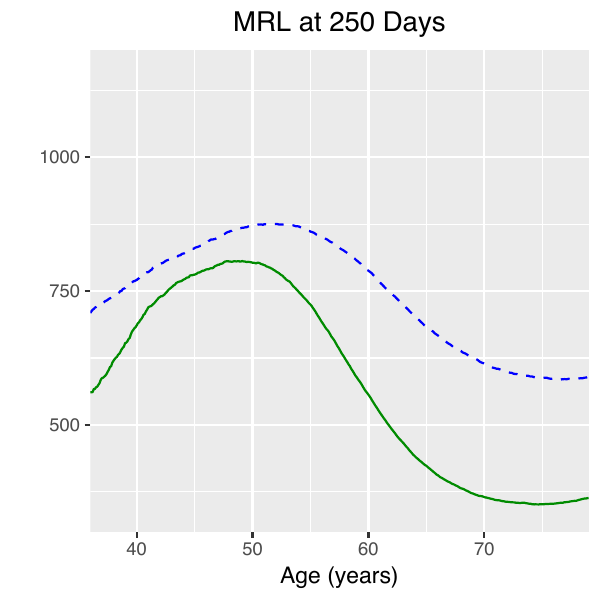}\includegraphics[height=1.8in,width=1.72in]{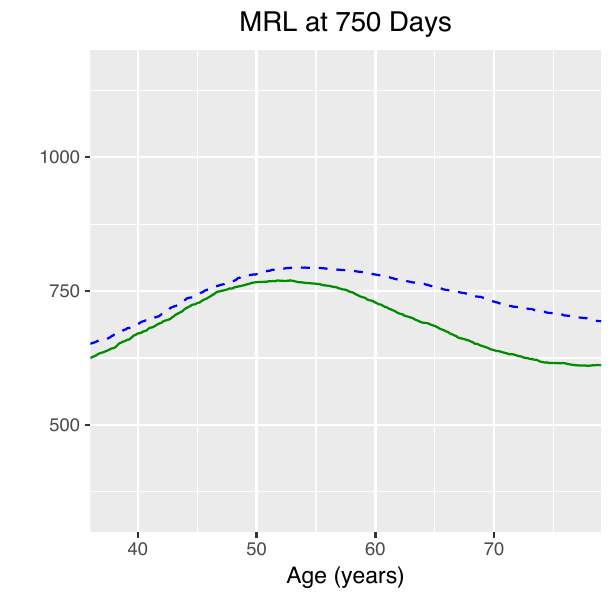} 
        	 \caption{Small cell lung cancer data.
Estimates of the MRL function of Arm A (blue dashed) and 
Arm B (green solid) for fixed ages (top panel) and for fixed times (bottom panels)}\label{fig:MRLreg}
\end{figure}

We also look at the MRL regression function at age $50$, $60$, and
$78$, see top panels in Figure \ref{fig:MRLreg}.  At age $50$, the MRL
function for Arm A appear monotonic while the MRL of Arm B has a very
shallow dip at about $400$ days then becomes indistinguishable from
Arm A.  At age $60$, the separation becomes more apparent towards in
the earlier survival range, and the dips are more pronounced and
present in both groups.  At age $78$, we see a similar curvature as in
our past analysis: a dip around $300-400$ and a shallow mode around
$1000-1200$.  While the shapes and range of the MRL functions change
across the covariate space, Arm A remains as high or higher than Arm B.

In the bottom panels of Figure \ref{fig:MRLreg}, we consider the MRL
as a function of age for three fixed time points: $0$, $250$, and
$750$ days. Recall that the mean regression function is equivalent to
the MRL at time $0$.  Therefore, the bottom left panel is simply the
mean regression function estimates (as in Figure \ref{fig:condmean})
for the two groups. At $250$ and $750$ days, we see a global decrease
in the remaining life expectancy compared to time $0$. At all times,
the maximum remaining life expectancy occurs around age $52$ years for
both groups.  The differentiation between groups is apparent across
all ages at $0$ and $250$ days, but is much less at $750$ days.  As
seen previously, Arm A appears to have a higher MRL across all ages at
all three times.  Moreover, the shape of the MRL as a function of age is non-linear and non-monotonic.

%
%

\section{Summary}\label{sec:conc}

We have proposed a nonparametric mixture model for mean
residual life (MRL) regression, a problem that, to our knowledge, has not
received attention in the Bayesian literature (parametric or nonparametric). 
The focus has been on developing general inference methodology for
both MRL functions across different values in the covariate space and
for MRL regression relationships across different time points. The
modeling approach builds from Dirichlet process mixture density
regression, including dependent Dirichlet process priors to
accommodate data from different experimental groups. The methodology
has been illustrated with both synthetic and real data examples.


\vspace{0.9cm}

\begin{center} 
{\Large \textbf{APPENDIX A: Theoretical Results}} 
\end{center}  

\noindent
{\bf Proof of the Lemma.}\\
Based on the DP constructive definition, 
$$
\text{E}(T \mid {\boldsymbol x},G) =
\sum_{l = 1}^{\infty} q_l({\boldsymbol x}; {\boldsymbol \theta}_{1l}) 
\text{E}(T \mid {\boldsymbol x},{\boldsymbol \theta}_{2l}) =
\frac{\sum_{l = 1}^{\infty} w_{l} A_{{\boldsymbol x}}({\boldsymbol \theta}_{l})}
{f({\boldsymbol x} \mid G)}
$$
where $A_{{\boldsymbol x}}({\boldsymbol \theta}) =$
$\int_{\mathbb{R}^{+}} u \, k(u, {\boldsymbol x} \mid {\boldsymbol \theta}) \, \text{d}u =$
$k({\boldsymbol x} \mid {\boldsymbol \theta}_{1}) 
\text{E}(T \mid {\boldsymbol x},{\boldsymbol \theta}_{2})$ $< \infty$, from the first lemma assumption. 
Let $Z_{{\boldsymbol x}} =$ $\sum_{l = 1}^{\infty} w_{l} A_{{\boldsymbol x}}({\boldsymbol \theta}_{l})$. 
Using the monotone convergence theorem, and the independence between the DP atoms and weights, 
we have $\text{E}(Z_{{\boldsymbol x}}) =$ 
$\sum_{l = 1}^{\infty} \text{E}(w_{l}) \text{E}(A_{{\boldsymbol x}}({\boldsymbol \theta}_{l})) =$
$\text{E}(A_{{\boldsymbol x}}({\boldsymbol \theta}_{l}))$, since this expectation is free of $l$ as the 
${\boldsymbol \theta}_{l}$ are i.i.d. (from $G_{0}$). Moreover, 
$$
\text{E}(A_{{\boldsymbol x}}({\boldsymbol \theta}_{l})) = \int A_{{\boldsymbol x}}({\boldsymbol \theta}) \,
\text{d}G_{0}({\boldsymbol \theta}_{1},{\boldsymbol \theta}_{2}) =
\left\{ \int k({\boldsymbol x} \mid {\boldsymbol \theta}_{1}) \, \text{d} G_{10}({\boldsymbol \theta}_{1}) \right\}
\left\{ \int \text{E}(T \mid {\boldsymbol x},{\boldsymbol \theta}_{2}) \,
\text{d}G_{20}({\boldsymbol \theta}_{2}) \right\}
$$ 
which is finite based on the second lemma assumption.
Since $Z_{{\boldsymbol x}}$ is a positive-valued random variable with finite expectation, we conclude
that $Z_{{\boldsymbol x}} < \infty$, almost surely, and therefore, 
$\text{E}(T \mid {\boldsymbol x},G) < \infty$, almost surely.

\vspace{0.5cm}

\noindent
{\bf Correlations of the DDP mixture prior model.}\\
The bivariate beta distribution by \citet{kotz} is based off of the product of independent beta distributions.  Start with sampling the independent latent variables: $U\sim \text{Beta}(\alpha, 1-b)$, $V\sim \text{Beta}(\alpha, 1-b)$, $W\sim \text{Beta}(\alpha +1 -b, b)$.  Let $\zeta_C = UW$ and $\zeta_T = VW$.  The weights are defined by $w_{1s}=1-\zeta_{1s}$, $w_{ls}=(1-\zeta_{ls})\prod_{r=1}^{l-1}\zeta_{rs}$, for $l\in\{2,3,...\}$. 

We are interested in obtaining the correlation between the two mixing
distributions, $G_C$ and $G_T$, implied under this bivariate beta
distribution.  We first start with the correlation between $\zeta_{C}$ and $\zeta_{T}$, $\text{Corr}(\zeta_{C},\zeta_{T})$.  We omit the component subscript in the latent variables, since results are the same for each $l\in \{1,2,...\}$. The covariance can be written as, $\text{Cov}(\zeta_{C},\zeta_{T}) = E(\zeta_{C}\zeta_{T}) - E(\zeta_{C})E(\zeta_{T}) = E((UW)(VW)) - E(UW)E(VW)$.  Using the fact that $U,V,W$ are independent, the covariance becomes $E(U)E(V)E(W^2) - E(U)E(V)E^2(W) = E(U)E(V)Var(W)$. Now, since $\zeta_C$ and $\zeta_T$ have the same marginal distribution, $\text{Beta}(\alpha,1)$,  the covariance and correlation reduces to:
{\small \begin{eqnarray*} \text{Cov}(\zeta_{C},\zeta_{T})&=& 
 \frac{\alpha^2b}{(\alpha +1 - b)(\alpha +1)^2(\alpha +2)} \nonumber \\
  \text{Corr}(\zeta_{C},\zeta_{T})&=& \frac{\alpha b}{\alpha+1-b} 
\end{eqnarray*}}
The correlation between $\zeta_C$ and $\zeta_T$ can take values on the interval $(0,1)$.  As $b\to0$ and/or $\alpha\to 0$, the correlation goes to $0$.  As $b\to 1$ and/or $\alpha\to \infty$, the correlation tends to $1$.

The next step is to explore the correlation of the weights, $\text{Corr}(w_{lC},w_{lT})$ for $l\in\{1,2,....\}$.  When $l=1$, $w_{1s} = 1-\zeta_{1s}$, which is simply a linear operation, hence the covariance and correlation are the same as before.  The $\text{Cov}(w_{1C},w_{1T}) = \text{Cov}(\zeta_{C},\zeta_{T})$ and $\text{Corr}(w_{1C},w_{1T}) = \text{Corr}(\zeta_{C},\zeta_{T})$ are given above.  The case is different for $l=\{2,3,...\}$.  In this case, the covariance is defined as$ E[((1-\zeta_{lC})\prod_{r=1}^{l-1}\zeta_{rC})((1-\zeta_{lT})\prod_{r=1}^{l-1}\zeta_{rT})] - E[ (1-\zeta_{lC})\prod_{r=1}^{l-1}\zeta_{rC}]E[(1-\zeta_{lT})\prod_{r=1}^{l-1}\zeta_{rT}]$.  Using the fact that $\zeta_{ls}$ are independent across $l=1,...,L$, for each $s\in\{C,T\}$, the covariance, for $l\in\{2,3,...\}$, can be expressed as
{\small\begin{eqnarray*}
 \text{Cov}(w_{lC},w_{lT})&=&\frac{(\alpha +1 -b)(\alpha+2)+\alpha^2b}{(\alpha +1 -b)(\alpha+1)^2(\alpha+2)}\left(\frac{\alpha^2b + \alpha^2(\alpha +1 -b)(\alpha+2)}{(\alpha +1 -b)(\alpha+1)^2(\alpha+2)}\right)^{l-1} \nonumber \\
 && - \frac{1}{(\alpha+1)^2}\left(\frac{\alpha^2}{(\alpha+1)^2}\right)^{l-1}
\end{eqnarray*} }
The variance for the weights are independent of group, and can be expressed as $Var(w_{ls}) = 2/(\alpha+1)(\alpha+2)[(\alpha+\alpha^2(\alpha+2))/((\alpha+1)^2(\alpha+2))]^{l-1} -1/(\alpha+1)^2[\alpha^2/(\alpha+1)^2]^{l-1}$. Therefore, the correlation, for $l\in\{2,3,...\}$, can be obtained by $\text{Corr}(w_{lC},w_{lT}) = \text{Cov}(w_{lC},w_{lT})/Var(w_{ls})$, which is in closed form, but does not reduce. The correlation between the weights for $l\in\{2,3,...\}$ also takes values on the interval $(0,1)$ and behaves the same in terms of the limits of $\alpha$ and $b$ as in the case when $l=1$. The component value, $l$, plays a slight role in the correlation, specifically as $l$ get larger, the rate of change for smaller $\alpha$ values becomes less extreme.    


The correlation between $G_T$ and $G_C$ is discussed in Section~\ref{sec:prop_ddpmm}.  Here, we provide details on the correlation between $T_C$ and $T_T$.  The $\text{Corr}(T_C, T_T)$ is found by marginalizing over the mixing distributions, $G_C$ and $G_T$.  Starting with the covariance, $\text{Cov}(T_C,T_T)=E[T_CT_T] - E[T_C]E[T_T] = E[E[T_C\mid G_C]E[T_T\mid G_T]] - E[E[T_C\mid G_C]]E[E[T_T\mid G_T]]$.  Under the gamma kernel with bivariate normal $G_0$ the covariance is given by the following,
{\small \begin{eqnarray*}
\text{Cov}(T_C,T_T) &=& \left(e^{t_2'{\boldsymbol \mu} +\frac{1}{2}t_2'{\boldsymbol \Sigma} t_2} - e^{2(t_3'{\boldsymbol \mu} +\frac{1}{2}t_3'{\boldsymbol \Sigma} t_3} \right) \left(\frac{(\alpha-2)b +\alpha +2}{\alpha(2\alpha -3b +5) -2b +2}\right)
\end{eqnarray*}}
where $t_2 = (2,-2)'$ and $t_3 = (1,-1)'$.  The variance of $T_s$, for both $s\in\{C,T\}$, is given by, $e^{t_1'{\boldsymbol \mu} +\frac{1}{2}t_1'{\boldsymbol \Sigma} t_1} +e^{t_2'{\boldsymbol \mu} +\frac{1}{2}t_2'{\boldsymbol \Sigma} t_2} - e^{2(t_3'{\boldsymbol \mu} +\frac{1}{2}t_3'{\boldsymbol \Sigma} t_3)}$.  Recall that $t_1=(1,-2)'$.  Therefore the correlation is given by,
{\small \begin{eqnarray*}
\text{Corr}(T_C,T_T) &=& \left[\left(e^{t_2'{\boldsymbol \mu} +\frac{1}{2}t_2'{\boldsymbol \Sigma} t_2} - e^{2(t_3'{\boldsymbol \mu} +\frac{1}{2}t_3'{\boldsymbol \Sigma} t_3)} \right)\left(\frac{(\alpha-2)b +\alpha +2}{\alpha(2\alpha -3b +5) -2b +2}\right)\right]\left. \middle/ \right. \nonumber  \\
&& \left[e^{t_1'{\boldsymbol \mu} +\frac{1}{2}t_1'{\boldsymbol \Sigma} t_1} +e^{t_2'{\boldsymbol \mu} +\frac{1}{2}t_2'{\boldsymbol \Sigma} t_2} - e^{2(t_3'{\boldsymbol \mu} +\frac{1}{2}t_3'{\boldsymbol \Sigma} t_3)}\right]    
\end{eqnarray*}}
As the $e^{t_1'{\boldsymbol \mu} +\frac{1}{2}t_1'{\boldsymbol \Sigma} t_1}= E[e^{\eta -2\phi}] \to 0$ the correlation simplifies to $((\alpha-2)b +\alpha +2)/(\alpha(2\alpha -3b +5) -2b +2)$.  In this case, as $\alpha \to 0$ the correlation tends to $1$ and as $\alpha \to \infty$ the correlation tends to $0$. Also, as $b \to 0$ the correlation tends to $1/(2\alpha +1)$ and as $b\to 1$ the correlation tends to $1/(\alpha +1)$.  These results are scaled down as $E[e^{\eta -2\phi}]$, the expectation of the kernel variance, gets larger.  


\begin{center} 
{\Large \textbf{APPENDIX B: MCMC Details}} 
\end{center}

%
%

\noindent
Here we show the posterior sampling algorithm used for the DDP mixture model in the presence of a single random continuous real-valued covariate as applied in Section 4.2 of the manuscript.  Omitting the covariate terms of the algorithm will yield the model applied in Sections 3.3 and 4.1. Assuming a singe group, i.e., $s$ having only a single index value, will yield the algorithm pertaining to the gamma DP mixture model applied in Section 2.3. 
We obtain posterior samples using the blocked Gibbs sampler and working with the latent parameters of the bivariate beta distribution. Posterior samples are based on a truncation approximation, $G_{Ls}$, to $G_s$:  $G_{Ls} = \sum_{l=1}^L p_{ls}\delta_{{\boldsymbol \theta}_l}$.  Specifically, the atoms are defined as ${\boldsymbol \theta}_l = (\eta_l, \phi_l, \beta_l, \kappa^2_l) \stackrel{\text{i.i.d.}}{\sim} G_0$, for $l=1,...,L$ with corresponding weights $p_{1s} = 1 - \zeta_{1s}, \ p_{ls} = (1- \zeta_{ls})\prod_{r=1}^{l-1} \zeta_{rs}$ for  $l\in\{2,3,...,L-1\}$ with $(\zeta_{lC}, \zeta_{lT})| {\boldsymbol \phi} \stackrel{\text{ind}}{\sim} \mbox{Biv-Beta}(\cdot \mid {\boldsymbol \phi})$, and $p_{Ls} = 1- \sum_{l=1}^{L-1}p_{ls}$.

Upon introducing the latent configuration variables, $\mathbf{w} = \{\mathrm{w}_{is} : i=1,...,n_s \mid s=C,T \}$, such that $w_{is} = l$ if the $i^{th}$ observation at group $s$ is assigned to mixture component $l$, the full hierarchical version of the model is written as,
\begin{eqnarray*}
(t_{is}, x_{is}) \mid {\mathrm w}_{is}, {\boldsymbol \theta}_l &\stackrel{\text{ind}}{\sim}& \Gamma(t_{is}\mid e^{\eta_{{\mathrm w}_{is}}}, e^{\phi_{{\mathrm w}_{is}}})\text{N}(x_{is}\mid \beta_{{\mathrm w}_{is}}, \kappa^2_{{\mathrm w}_{is}})\\
{\mathrm w}_{is} \mid \{(\zeta_{ls})\} &\stackrel{\text{ind}}{\sim}& \sum_{l=1}^L\{(1- \zeta_{ls}) \prod_{r=1}^{l-1}\zeta_{rs}\}\delta_l({\mathrm w}_{is}), \ \ \ \mbox{for} \  i=1,...,n_s \  \mbox{and} \ s\in\{C,T\}\\
\{(\zeta_{lC},\zeta_{lT})\}\mid \alpha, b &\sim& \text{Biv-Beta}(\{(\zeta_{lC},\zeta_{lT})\}\mid \alpha,b)\\
({\eta}_l, \phi_l)' \mid {\boldsymbol \mu},{\boldsymbol \Sigma}  & \stackrel{\text{i.i.d.}}{\sim}& \text{N}_2(({\eta}_l, \phi_l)' \mid {\boldsymbol \mu},{\boldsymbol \Sigma}), \ \ \mbox{for} \ l=1,...,L  
\end{eqnarray*}

\noindent where $\zeta_{lC} = UW$ and $\zeta_{lT} = VW$ for $l\in\{1,...,L\}$ with $U \stackrel{\text{i.i.d.}}{\sim} \text{Beta}(\alpha, 1-b), \ V \stackrel{\text{i.i.d.}}{\sim} \text{Beta}(\alpha, 1-b)$, and  $W  \stackrel{\text{i.i.d.}}{\sim} \text{Beta}(1+\alpha - b, b)$. We place the following priors: $\alpha \sim \Gamma(\alpha \mid a_\alpha, b_\alpha), \ b \sim \text{Unif}(b\mid 0,1), \ {\boldsymbol \mu} \sim \text{N}_2(\boldsymbol{\mu}\mid a_\mu, B_\mu), \  {\boldsymbol \Sigma} \sim  \text{IWish}({\boldsymbol \Sigma}\mid a_\Sigma, B_\Sigma),  \ \beta_l\mid \lambda,\tau^2  \stackrel{\text{i.i.d.}}{\sim} \text{N}(\beta_l \mid \lambda, \tau^2), \ \kappa^2_l \mid a, \ \rho  \stackrel{\text{i.i.d.}}{\sim}\Gamma^{-1}(\kappa^2 \mid a, \rho), \ \lambda \sim \text{N}(\lambda\mid a_\lambda, b^2_\lambda), \  \tau^2 \sim \Gamma^{-1}(\tau^2\mid a_\tau, b_\tau)$, and $\rho \sim \Gamma(\rho\mid a_\rho, b_\rho)$, with $l\in\{1,...,L\}$. 

Let $L_s^{*}$  be the number of distinct components, and ${\mathbf w}_{s}^*\equiv \{{\mathrm w}^*_{js}: j=1,..., L^*_s\}$ be the vector of latent configuration variables for group $s\in\{C,T\}$.  For subject $i =1,...,n_s$, let $\delta_{is} = 0$ if $t_{is}$ is observed and $\delta_{is} = 1$ if $t_{is}$ is right censored. Let $\Psi$ represent the vector of the most recent iteration of all other parameters.  Let $b=1,...,B$ be the number of iterations in the MCMC.  The posterior samples of $p({\boldsymbol \eta}, {\boldsymbol \phi}, {\boldsymbol \beta}, {\boldsymbol \kappa}^2,{\mathbf w},{\boldsymbol \zeta},{\boldsymbol \mu},{\boldsymbol \Sigma}, \lambda, \tau^2, \rho, \alpha, b\mid data)$ can be obtained by the following:

First, we consider updates for $({\eta}_l,\phi_l)'$,$\beta_l$, and $\kappa^2_l$ for $l= 1,..., L $. If $l$ is not already a component:  $ l \notin  {\mathbf w}_C^{*(b)} \cup {\mathbf w}_T^{*(b)} $, then draw $p({\eta}^{(b+1)}_l, \phi^{(b+1)}_l\mid data, \Psi)  \stackrel{\mbox{}}{\sim} N_2({\boldsymbol \mu}^{(b)},{\boldsymbol \Sigma}^{(b)})$, $p(\beta_l^{(b+1)}\mid data, \Psi) \stackrel{\mbox{}}{\sim} N(\lambda^{(b)}, \kappa_l^{2(b)})$, and  $p(\kappa_l^{2(b+1)} \mid data, \Psi) \stackrel{\mbox{}}{\sim} \Gamma^{-1}(a, \rho^{(b)})$.
If $l$ is an active component in either or both:  $ l \in  {\mathbf w}_C^{*(b)} \cup l \in  {\mathbf w}_T^{*(b)}$.  We have $p(\eta_l, \phi_l\mid data, { \Psi}) \propto$ $ N_2((\eta_l, \phi_l)' \mid {\boldsymbol \mu}, {\boldsymbol \Sigma})\prod_{s\in\{C,T\}}\prod_{\{i:l={\mathrm w}_{is}\}}[\Gamma(t_{is}\mid e^{\eta_l}, e^{\phi_l})]^{1-\delta_{is}}[\int_{t_{is}}^\infty \Gamma(u_i\mid e^{\eta_l}, e^{\phi_l})dt_i]^{\delta_{is}}$.  We use a Metropolis-Hastings step for this update.  We sample from the proposal distribution $(\eta_l', \phi_l')' \sim N_2((\eta_l^{(b)}, \phi_l^{(b)})', cS^2)$, where $S^2$ is updated from the average posterior samples of  ${\boldsymbol \Sigma}$ under initial runs, and $c>1$. 
For $\beta_l$ and $\kappa_l$, we have $p(\beta_l \mid data, { \Psi}) \propto  N(\beta_l \mid \lambda, \tau^2) \prod_{s\in\{C,T\}}\prod_{\{i:l={\mathrm w}_{is}\}}N(x_{is}\mid \beta_l, \kappa^2_l)$ and $p(\kappa_l^2\mid data, \Psi) \propto \Gamma^{-1}(\kappa_l^2\mid a, \rho)  \prod_{s\in\{C,T\}} $ $\prod_{\{i:l={\mathrm w}_{is}\}}N(x_{is}\mid \beta_l, \kappa^2_l)$. Thus we sample via:\\
{\small\begin{eqnarray*}
p(\beta_l^{(b+1)}\mid data, { \Psi}) \stackrel{\mbox{}}{\sim} N(m_\beta, s^2_\beta) \hspace{4.1in} \\
p(\kappa_l^{2(b+1)} \mid data, \Psi) \stackrel{\mbox{}}{\sim} \Gamma^{-1} \left(a + 0.5\sum_{s\in\{C,T\}}\sum_{\{i:l={\mathrm w}_{is}\}} 1,  \rho^{(b)} + 0.5\sum_{s\in\{C,T\}}\sum_{\{i:l={\mathrm w}_{is}\}} (x_{is} - \beta_l^{(b+1)})^2 \right) \hspace{0.5in}
\end{eqnarray*}}
 
\noindent where $m_\beta = s^2_\beta \left(\kappa_l^{-2(b)} \left[\sum_{s\in\{C,T\}}\sum_{\{i:l={\mathrm w}_{is}\}} x_{is}\right] + \tau^{-2(b)}\lambda^{(b)}\right)$, \\ and $s^2_\beta = \left(\tau^{-2(b)} + \kappa_l^{-2(b)}\left[\sum_{s\in\{C,T\}}\sum_{\{i:l={\mathrm w}_{is}\}} 1\right]\right)^{-1}$.

\vspace{.2in}

To obtain samples from $p({\boldsymbol \zeta}\mid {\Psi}.data)$ we work with $\{U_l,V_l,W_l\}$.  
Using slice sampling, we can introduce latent variables $\nu_l$ and $\gamma_l$ for $l=1,...,L$, such that we have Gibbs steps for each parameter:
{\small\begin{eqnarray*}
&&p(\nu_l^{(b+1)}\mid {\Psi},data) \sim \text{Unif}\left(0,(1-U^{(b)}_lW^{(b)}_l)^{M^{(b)}_{lC}} \right)\\
&&p(\gamma_l^{(b+1)}\mid {\Psi},data) \sim \text{Unif}\left(0,(1-V^{(b)}_l W^{(b)}_l)^{M^{(b)}_{lT}}\right) \\
&&p(U_l^{(b+1)}\mid \Psi, data) \sim \text{Beta}\left((\sum_{r=l+1}^L M^{(b)}_{rC}) +\alpha, 1-b\right){\boldsymbol 1}_{\left(0, \frac{1}{W^{(b)}_l}\left[1- exp\left(\frac{log(\nu^{(b+1)}_l)}{M^{(b)}_{lC}}\right)\right]\right)}\\
&&p(V_l^{(b+1)}\mid \Psi, data) \sim \text{Beta}\left((\sum_{r=l+1}^L M^{(b)}_{rT}) +\alpha, 1-b\right){\boldsymbol 1}_{\left(0, \frac{1}{W^{(b)}_l}\left[1- exp\left(\frac{log(\gamma^{(b+1)}_l)}{M^{(b)}_{lT}}\right)\right]\right)}\\ 
&&p(W_l^{(b+1)}\mid \Psi, data) \sim \text{Beta}\left((\sum_{r=l+1}^L  M^{(b)}_{rT}+ M^{(b)}_{rC}) +\alpha +1 -b, b\right){\boldsymbol 1}_{\left(0,m^*\right)}
\end{eqnarray*}}

\noindent where  $m^* = \text{min}\left\{ \frac{1}{U^{(b+1)}_l}\left[1- exp\left(\frac{log(\nu^{(b+1)}_l)}{M^{(b)}_{lC}}\right)\right], \frac{1}{V^{(b+1)}_l}\left[1- exp\left(\frac{log(\gamma^{(b+1)}_l)}{M^{(b)}_{lT}}\right)\right]\right\}\\
\mbox{Set} \ \zeta^{(b+1)}_{lC} = U^{(b+1)}_lW^{(b+1)}_l$ and $\zeta^{(b+1)}_{lT} = V^{(b+1)}_lW^{(b+1)}_l 
$\\

For the update for ${\mathrm w}_{is}$ we have $p({\mathrm w}_{is}\mid data,\Psi) \propto  \Gamma(t_{is}\mid e^{\eta_{{\mathrm w}_{is}}}, e^{\phi_{{\mathrm w}_{is}}})\text{N}(x_{is}\mid \beta_{{\mathrm w}_{is}}, \kappa^2_{{\mathrm w}_{is}})  \sum_{l=1}^L\{(1- \zeta_{ls}) \prod_{r=1}^{l-1}\zeta_{rs}\}\delta_l({\mathrm w}_{is})$, so we sample from  $p({\mathrm w}^{(b+1)}_{is}\mid data,\Psi)  \stackrel{}{\sim}  \sum_{l=1}^L\tilde{p}_{lis}\delta_{(l)}({\mathrm w}_{is})$ 
where $\tilde{p}_{lis} = p_{ls}[\Gamma(t_{is}\mid e^{\eta_l^{(b+1)}}, e^{\phi_l^{(b+1)}})]^{1-\delta_{is}} [\int_{t_{is}}^\infty\Gamma(u_{is}\mid e^{\eta_l^{(b+1)}}, e^{\phi_l^{(b+1)}})du_{is}]^{\delta_{is}}\text{N}(x_{is}\mid \beta^{(b+1)}_l, \kappa^{2(b+1)}_l)/\{\sum_{l=1}^Lp_{ls}[\Gamma(t_{is}\mid e^{\eta_l^{(b+1)}}, e^{\phi_l^{(b+1)}})]^{1-\delta_{is}}[\int_{t_{is}}^\infty\Gamma(u_{is}\mid e^{\eta_l^{(b+1)}}, e^{\phi_l^{(b+1)}})du_{is}]^{\delta_{is}}\text{N}(x_{is}\mid  \beta^{(b+1)}_l, \kappa^{2(b+1)}_l)\}$ with $p_{1s} = 1-\zeta_{1s}$ and $p_{ls}=(1-\zeta_{ls})\prod_{r=1}^{l-1}\zeta_{rs}$ for $l = 2,..., L-1$.\\

For the update for ${\boldsymbol \mu} $ we have $p({\boldsymbol \mu}  \mid data, \Psi) \propto \text{N}_2({\boldsymbol \mu}\mid a_\mu, B_\mu)\prod_{l=1}^L \text{N}_2((\eta_l, \phi_l)'\mid {\boldsymbol \mu}, {\boldsymbol \Sigma})$, so we sample $p({\boldsymbol \mu}^{(b)}  \mid data, \Psi) \stackrel{\mbox{}}{\sim}  \text{N}_2(m_\mu, S_\mu^2)$ where $m_\mu = S_\mu^2(B_\mu^{-1}a_\mu + {\boldsymbol \Sigma}^{-1}\sum_{l=1}^L(\eta_l, \phi_l)'^{(b)})$, $ S_\mu^2 = (B_\mu^{-1} + L{\boldsymbol \Sigma}^{-1(b)})^{-1}$.  \\

For the update of ${\boldsymbol \Sigma}$, we have $p({\boldsymbol \Sigma} \mid data, \Psi) \propto \prod_{l=1}^L \text{N}_2((\eta_l, \phi_l)'\mid {\boldsymbol \mu}, {\boldsymbol \Sigma}) \text{IWish}({\boldsymbol \Sigma} \mid a_\Sigma, B_\Sigma)$, so we sample $p({\boldsymbol \Sigma}^{(b+1)} \mid data, \Psi) \stackrel{\mbox{}}{\sim} \text{IWish}(L + a_\Sigma, B_\Sigma + \sum_{l=1}^L ((\eta_l, \phi_l)'^{(b+1)} - {\boldsymbol \mu}^{(b+1)})((\eta_l, \phi_l)'^{(b+1)} - {\boldsymbol \mu}^{(b+1)})')$\\

For the update for $\lambda$ we have $p(\lambda \mid data, \Psi) \propto  \text{N}(\lambda \mid a_\lambda, b^2_\lambda)\prod_{l=1}^L \text{N}(\beta_l \mid \lambda, \tau^2)$, so we sample $p(\lambda^{(b+1)}\mid data, \Psi)  \stackrel{\mbox{}}{\sim} \text{N}(m_\lambda, s^2_\lambda)$ where $m_\lambda = s^2_\lambda(b_\lambda^{-2}a_\lambda + \tau^{-2}\sum_{l=1}^L \beta_l)$ and $s^2_\lambda = ( b^{-2}_\lambda + \tau^{-2(b)}L)^{-1}$. \\

For the update for $\tau^2$ we have $p(\tau^2 \mid data, \Psi) \propto  \Gamma^{-1}(\tau^2\mid a_\tau, b_\tau)\prod_{l=1}^L N(\beta_l \mid \lambda, \tau^2)$, so we sample $p(\tau^{2(b+1)} \mid data, \Psi)  \stackrel{\mbox{}}{\sim} \Gamma^{-1}(0.5L + a_\tau, 0.5[\sum_{l=1}^L (\beta_l^{(b+1)} - \lambda^{(b+1)})^2 ] + b^\tau)$ \\

For the update for $\rho$, $p(\rho \mid data, \Psi) \propto \Gamma(\rho\mid  a_\rho,b_\rho)\prod_{l=1}^L  \Gamma^{-1}(\kappa^2_l\mid a, \rho)$, so we sample $p(\rho^{(b+1)} \mid data, \Psi) \stackrel{\mbox{}}{\sim} \Gamma(aL+ a_\rho, [\sum_{l=1}^L  \kappa_l^{-2(b+1)}] + b_\rho)$. \\

We do not have conjugacy for $\alpha$ and $b$, so we turn to the Metropolis-Hastings algorithm to update these parameters.  The Bivariate Beta density of $(\zeta_{c}, \zeta_{T})$, has a complicated form, however, we can work with the density of the latent variables, $(U,V,W)$: $p(\alpha, b \mid data, \Psi) \propto  \text{Unif}(b\mid 0,1)\Gamma(\alpha\mid  a_\alpha, b_\alpha)\prod_{l-1}^{L-1}\text{Beta}(U_l\mid \alpha,1-b)\text{Beta}(V_l\mid \alpha, 1-b) \text{Beta}(W_l\mid 1+\alpha - b, b)$. We sample from the proposal distribution,  $(\log(\alpha'), \text{logit}(b'))' \sim \text{N}_2((\log(\alpha^{(b)}),\text{logit}(b^{(b)})), cS^2_{\alpha b})$, where $S_{\alpha b}^2 $ is updated from the average variances and covariance of posterior samples of $((\log(\alpha),\text{logit}(b))$ under initial runs, and $c $ is updated from initial runs to optimize mixing.  \\

\begin{center} 
{\Large \textbf{APPENDIX C: Conditional Predictive Ordinate Derivations}} 
\end{center}

\noindent
Here we provide the details of how we arrived to the expression necessary for computing the CPO values under the DDP mixture model.  As our data example in Section 4.1 does not contain any random covariates, we will derive the expression without covariates, however, the derivation can easily be extended to include random covariates in the curve-fitting setting.  The hierarchical form of the  DDP mixture model without covariates and based on the truncation approximation, $G_{Ls}$, of $G_s$  is given as follows:
\begin{eqnarray*}
t_{is}|{\mathrm w}_{is}, {\boldsymbol \theta} &\stackrel{\text{ind}}{\sim}& \Gamma(t_{is}\mid   {\boldsymbol \theta}_{{\mathrm w}_{is}} )\  \ \mbox{for} \ i= 1,...,n_s \ \  s\in\{C,T\} \\
 {\boldsymbol {\mathrm w}}\mid \{\zeta_{lC}, \zeta_{lT}\} &\sim& \prod_{s\in \{C,T\}}\prod_{i=1}^{n_s}\sum_{l=1}^L\left[(1-\zeta_{ls})\prod_{r=1}^{l-1}\zeta_{rs}\right]\delta_l ({\mathrm w}_{is}) \\
 {\boldsymbol \theta}_l \mid {\boldsymbol \mu}, {\boldsymbol \Sigma} &\stackrel{\text{i.i.d.}}{\sim}& \text{N}_2( {\boldsymbol \theta}_l\mid {\boldsymbol \mu}, {\boldsymbol \Sigma}) \\
 (\zeta_{lC},\zeta_{lT})\mid \alpha , b &\stackrel{\text{i.i.d.}}{\sim}& \text{Biv-Beta}( (\zeta_{lC},\zeta_{lT})\mid \alpha , b) \  \ \mbox{for} l= 1,...,L-1 
 \end{eqnarray*}
with $\alpha \sim \Gamma(\alpha\mid a_\alpha,b_\alpha)$, $b\sim \text{Unif}(b\mid 0,1)$, ${\boldsymbol \mu}\sim \text{N}_2({\boldsymbol \mu}\mid a_\mu, B_\mu)$, and ${\boldsymbol \Sigma}\sim \text{IWish}({\boldsymbol \Sigma}\mid a_\Sigma, B_\Sigma)$.  Let $\Psi = (\alpha, b, {\boldsymbol \mu}, {\boldsymbol \Sigma})$. The predictive density for a new survival time from group $s$, $t_{0s}$, is given by:
\begin{eqnarray*}
p(t_{0s}\mid data) &=& \int\int\Gamma(t_{0s}\mid  {\boldsymbol \theta}_{{\mathrm w}_{0s}})\left(\sum_{l=1}^Lp_{ls}\delta_l({\mathrm w}_{0s})\right)p({\boldsymbol \theta},{\boldsymbol p}, {\boldsymbol {\mathrm w}}, \Psi \mid data)d{\mathrm w}_{0s}d{\boldsymbol \theta}d{\boldsymbol {\mathrm w}} d{\boldsymbol p} d\Psi \\
 &=& \int \left(\sum_{l=1}^L p_{ls}\Gamma(t_{0s}\mid {\boldsymbol \theta}_l)\right)p({\boldsymbol \theta},{\boldsymbol p}, {\boldsymbol {\mathrm w}}, \Psi \mid data)d{\boldsymbol \theta}d{\boldsymbol {\mathrm w}} d{\boldsymbol p} d\Psi
\end{eqnarray*}
\indent Let $s'$ be the experimental group that $s$ is not, $data = \{{\boldsymbol t}_s, {\boldsymbol t}_{s'}\}$, and $A$ be the normalizing constant for $p({\boldsymbol \theta},{\boldsymbol p}, {\boldsymbol {\mathrm w}}, \Psi \mid data)$.  Namely, $p({\boldsymbol \theta},{\boldsymbol p}, {\boldsymbol {\mathrm w}}, \Psi \mid data) = [(\prod_{i=1}^{n_s} \Gamma(t_{is}\mid {\boldsymbol \theta}_{{\mathrm w}_{is}})) $ $ (\prod_{i=1}^{n_{s'}} \Gamma(t_{is'}\mid {\boldsymbol \theta}_{{\mathrm w}_{is'}}))p({\boldsymbol \theta},{\boldsymbol p}, {\boldsymbol {\mathrm w}}, \Psi)] / [\int (\prod_{i=1}^{n_s} \Gamma(t_{is}\mid {\boldsymbol \theta}_{{\mathrm w}_{is}}))(\prod_{i=1}^{n_{s'}} \Gamma(t_{is'}\mid  {\boldsymbol \theta}_{{\mathrm w}_{is'}}))p({\boldsymbol \theta},{\boldsymbol p}, {\boldsymbol {\mathrm w}}, \Psi) d{\boldsymbol \theta}$\\$d{\boldsymbol {\mathrm w}} d{\boldsymbol p} d\Psi]$. Note that $p({\boldsymbol \theta},{\boldsymbol p}, {\boldsymbol {\mathrm w}}, \Psi) = \text{N}_2({\boldsymbol \theta}\mid  {\boldsymbol \mu}, {\boldsymbol \Sigma})(\prod_{i=1}^{n_{s}}\sum_{l=1}^L p_{ls} \delta_l({\mathrm w}_{is})) (\prod_{i=1}^{n_{s'}}\sum_{l=1}^L p_{ls'} $ $\delta_l({\mathrm w}_{is'}))  \text{Biv-Beta}({\boldsymbol p}\equiv (\zeta_s,\zeta_{s'})\mid  \alpha, b) \Gamma(\alpha\mid a_\alpha, b_\alpha)\text{Unif}(b\mid 0,1)N_2({\boldsymbol \mu}\mid a_\mu,B_\mu) \text{IWish}({\boldsymbol \Sigma}\mid a_\Sigma, $ $B_\Sigma)$. 

The CPO of the $ith$ survival time in group $s$ is defined as, $\text{CPO}_{is} = p(t_{is}\mid {\boldsymbol t}_{(-i)s}, {\boldsymbol t}_{s'}) = \int \Gamma(t_{is}\mid {\boldsymbol \theta}_{{\mathrm w}_{0s}})(\sum_{l=1}^L p_{ls}\delta_l({\mathrm w}_{0s}))p({\boldsymbol \theta},{\boldsymbol p}, {\boldsymbol {\mathrm w}}_{(-i)s}, \Psi) d{\boldsymbol \theta}d{\boldsymbol {\mathrm w}_{(-i)s}} d{\boldsymbol p} d\Psi d{\mathrm w}_{0s}$, where ${\boldsymbol {\mathrm w}}_{(-i)s}$ is the vector ${\boldsymbol {\mathrm w}}$ with the $i^{th}$ member of group $s$ removed. Similarly, $data_{(-i)s}$ represents $data$ with the $i^{th}$ member in group $s$ removed.  Now, consider $p({\boldsymbol \theta}, {\boldsymbol p}, {\boldsymbol {\mathrm w}}_{(-i)s}, \Psi|data_{(-i)s})$, which is given by:
\begin{eqnarray*}
&& \frac{p(data_{(-i)s}\mid {\boldsymbol \theta},{\boldsymbol {\mathrm w}}_{(-i)s}) p({\boldsymbol \theta}, {\boldsymbol {\mathrm w}}_{(-i)s}, {\boldsymbol p},\Psi)}{\int p(data_{(-i)s}\mid {\boldsymbol \theta}, {\boldsymbol {\mathrm w}}_{(-i)s}) p({\boldsymbol \theta}, {\boldsymbol {\mathrm w}}_{(-i)s}, {\boldsymbol p},\Psi)d{\boldsymbol {\mathrm w}_{(-i)s}}  d{\boldsymbol p} d\Psi} \\
&&= \frac{\left\{\prod_{j\neq i}^{n_s} \Gamma(t_{js}\mid {\boldsymbol \theta}_{{\mathrm w}_{js}})\right\}\left\{\prod_{i=1}^{n_{s'}} \Gamma(t_{is'}\mid {\boldsymbol \theta}_{{\mathrm w}_{is'}})\right\}p({\boldsymbol \theta},{\boldsymbol p}, {\boldsymbol {\mathrm w}}_{(-i)s}, \Psi)}{\int \left\{\prod_{j\neq i}^{n_s} \Gamma(t_{js}\mid {\boldsymbol \theta}_{{\mathrm w}_{js}})\right\}\left\{\prod_{i=1}^{n_{s'}} \Gamma(t_{is'}\mid {\boldsymbol \theta}_{{\mathrm w}_{is'}})\right\}p({\boldsymbol \theta},{\boldsymbol p}, {\boldsymbol {\mathrm w}}_{(-i)s}, \Psi) d{\boldsymbol \theta}d{\boldsymbol {\mathrm w}}_{(-i)s} d{\boldsymbol p} d\Psi}
\end{eqnarray*}
Let $B_{is}$ be the normalizing constant of $p({\boldsymbol \theta}, {\boldsymbol p}, {\boldsymbol {\mathrm w}}_{(-i)s}, \Psi\mid data_{(-i)s})$, specifically:\\ 
$B_{is} = \int \left\{\prod_{j\neq i}^{n_s} \Gamma(t_{js}\mid {\boldsymbol \theta}_{{\mathrm w}_{js}})\right\}\left\{\prod_{i=1}^{n_{s'}} \Gamma(t_{is'}\mid {\boldsymbol \theta}_{{\mathrm w}_{is'}})\right\}p({\boldsymbol \theta},{\boldsymbol p}, {\boldsymbol {\mathrm w}}_{(-i)s}, \Psi) d{\boldsymbol \theta}d{\boldsymbol {\mathrm w}}_{(-i)s} d{\boldsymbol p} d\Psi$\\
Then, we can write $p({\boldsymbol \theta}, {\boldsymbol p}, {\boldsymbol {\mathrm w}}_{(-i)s}, \Psi|data_{(-i)s}) $ as:
\begin{eqnarray*}
 &&\frac{\left\{\prod_{i=1}^{n_s} \Gamma(t_{is}\mid {\boldsymbol \theta}_{{\mathrm w}_{is}})\right\}\left\{\prod_{i=1}^{n_{s'}} \Gamma(t_{is'}\mid {\boldsymbol \theta}_{{\mathrm w}_{is'}})\right\}p({\boldsymbol \theta},{\boldsymbol p}, {\boldsymbol {\mathrm w}}, \Psi)}{B_{is} \Gamma(t_{is}\mid {\boldsymbol \theta}_{{\mathrm w}_{is}}) p({\mathrm w}_{is}\mid {\boldsymbol p})}=\frac{A}{B_{is}} \frac{ p({\boldsymbol \theta},{\boldsymbol p}, {\boldsymbol {\mathrm w}}, \Psi\mid data)}{\Gamma(t_{is} \mid {\boldsymbol \theta}_{{\mathrm w}_{is}}) p({\mathrm w}_{is}\mid{\boldsymbol p})}
\end{eqnarray*}
Thus, 
\begin{eqnarray*}
\text{CPO}_{is} &=& \int \Gamma(t_{is}\mid {\boldsymbol \theta}_{{\mathrm w}_{0s}})p({\mathrm w}_{0s}\,id {\boldsymbol p})p({\boldsymbol \theta},{\boldsymbol p}, {\boldsymbol {\mathrm w}}_{(-i)s}, \Psi) d{\boldsymbol \theta}d{\boldsymbol {\mathrm w}_{(-i)s}} d{\boldsymbol p} d\Psi d{\mathrm w}_{0s} \\
&=& \int \Gamma(t_{is}\mid {\boldsymbol \theta}_{{\mathrm w}_{0s}}) \left( \int p( {\mathrm w}_{0s},{\mathrm w}_{is}\mid {\boldsymbol p})d{\mathrm w}_{is} \right)p({\boldsymbol \theta},{\boldsymbol p}, {\boldsymbol {\mathrm w}}_{(-i)s}, \Psi) d{\boldsymbol \theta}d{\boldsymbol {\mathrm w}_{(-i)s}} d{\boldsymbol p} d\Psi d{\mathrm w}_{0s}\\
&=& \frac{A}{B_{is}}\frac{\int \left[\Gamma(t_{is} \mid {\boldsymbol \theta}_{{\mathrm w}_{0s}}) p({\mathrm w}_{0s},{\mathrm w}_{is}\mid {\boldsymbol p})\right]}{\left[\Gamma(t_{is} \mid {\boldsymbol \theta}_{{\mathrm w}_{is}}) p({\mathrm w}_{is}\mid {\boldsymbol p})\right]p({\boldsymbol \theta}, {\boldsymbol p}, {\boldsymbol {\mathrm w}}, \Psi\mid data) d{\mathrm w}_{0s}d{\boldsymbol \theta}d{\boldsymbol {\mathrm w}} d{\boldsymbol p} d\Psi }\\
&=& \frac{A}{B_{is}}\frac{ \int \left[\sum_{l=1}^L p_{ls} \Gamma(t_{is} \mid {\boldsymbol \theta}_l)\right]}{\left[\Gamma(t_{is}\mid  {\boldsymbol \theta}_{{\mathrm w}_{is}})\right] p({\boldsymbol \theta}, {\boldsymbol p}, {\boldsymbol {\mathrm w}}, \Psi\mid data) d{\mathrm w}_{0s}d{\boldsymbol \theta}d{\boldsymbol {\mathrm w}} d{\boldsymbol p} d\Psi}
\end{eqnarray*}

Note, $p({\mathrm w}_{0s}\mid {\mathrm w}_{is},{\boldsymbol p}) = p({\mathrm w}_{0s}\mid {\boldsymbol p}) $. All that is left is to be able to evaluate $A/B_{is}$:
\begin{eqnarray*}
\left(\frac{A}{B_{is}}\right)^{-1} &=& \frac{1}{A} \int \left\{\prod_{j\neq i}^{n_s}\Gamma(t_{js}\mid {\boldsymbol \theta}_{{\mathrm w}_{js}})\right\} \left\{\prod_{i=1}^{n_{s'}}\Gamma(t_{is'}\mid  {\boldsymbol \theta}_{{\mathrm w}_{is'}})\right\}\underbrace{\left(\int p({\mathrm w}_{is}\mid{\boldsymbol {\mathrm w}}_{(-i)s}, {\boldsymbol p})d{\mathrm w}_{is}\right)}_{1} \\
&& \times p({\boldsymbol {\mathrm w}}_{(-i)s} \mid {\boldsymbol p}) p({\boldsymbol p}, {\boldsymbol \theta},\Psi) d{\boldsymbol \theta}d{\boldsymbol {\mathrm w}}_{(-i)s} d{\boldsymbol p} d\Psi  \\
&=&  \frac{1}{A} \int \left\{\prod_{j\neq i}^{n_s}\Gamma(t_{js}\mid  {\boldsymbol \theta}_{{\mathrm w}_{js}})\right\} \left\{\prod_{i=1}^{n_{s'}}\Gamma(t_{is'}\mid {\boldsymbol \theta}_{{\mathrm w}_{is'}})\right\}p({\boldsymbol \theta}, {\boldsymbol p}, {\boldsymbol {\mathrm w}}, \Psi) d{\boldsymbol \theta}d{\boldsymbol {\mathrm w}} d{\boldsymbol p} d\Psi \\
&=&  \frac{1}{A} \int \frac{\left\{\prod_{j\neq i}^{n_s}\Gamma(t_{js}\mid {\boldsymbol \theta}_{{\mathrm w}_{js}})\right\} \left\{\prod_{i=1}^{n_{s'}}\Gamma(t_{is'}\mid  {\boldsymbol \theta}_{{\mathrm w}_{is'}})\right\}}{\Gamma(t_{is}\mid{\boldsymbol \theta}_{{\mathrm w}_{is}})}p({\boldsymbol \theta}, {\boldsymbol p}, {\boldsymbol {\mathrm w}}, \Psi) d{\boldsymbol \theta}d{\boldsymbol {\mathrm w}} d{\boldsymbol p} d\Psi\\
&=& \int \frac{1}{\Gamma(t_{is}\mid  {\boldsymbol \theta}_{{\mathrm w}_{is}})}p({\boldsymbol \theta}, {\boldsymbol p}, {\boldsymbol {\mathrm w}}, \Psi) d{\boldsymbol \theta}d{\boldsymbol {\mathrm w}} d{\boldsymbol p} d\Psi
\end{eqnarray*}
Collecting the final terms,
\begin{eqnarray*}
\text{CPO}_{is} &=& \frac{A}{B_{is}}\frac{ \int \left[\sum_{l=1}^L p_{ls} \Gamma(t_{is} \mid {\boldsymbol \theta}_l)\right]}{\left[\Gamma(t_{is}\mid  {\boldsymbol \theta}_{{\mathrm w}_{is}})\right] p({\boldsymbol \theta}, {\boldsymbol p}, {\boldsymbol {\mathrm w}}, \Psi| data) d{\mathrm w}_{0s}d{\boldsymbol \theta}d{\boldsymbol {\mathrm w}} d{\boldsymbol p} d\Psi} \\
&& \text{where} \ \left(\frac{A}{B_{is}}\right)^{-1} =  \int \frac{1}{\Gamma(t_{is}\mid {\boldsymbol \theta}_{{\mathrm w}_{is}})}p({\boldsymbol \theta}, {\boldsymbol p}, {\boldsymbol {\mathrm w}}, \Psi) d{\boldsymbol \theta}d{\boldsymbol {\mathrm w}} d{\boldsymbol p} d\Psi
\end{eqnarray*}
The MCMC approximation of the CPO values is given by:
\begin{eqnarray*}
\text{CPO}_{is} &\approx& \frac{A}{B_{is}}\left( \sum_{j=1}^B \frac{\sum_{l=1}^L \Gamma(t_{is} \mid {\boldsymbol \theta}_{lj})}{\Gamma(t_{is} \mid {\boldsymbol \theta}_{{\boldsymbol {\mathrm w}}_{isj}})}\right),  \mbox{ \ \ where \ \ }  \frac{A}{B_{is}} = \left( \sum_{j=1}^B \frac{1}{\Gamma(t_{is} \mid {\boldsymbol \theta}_{{\boldsymbol {\mathrm w}}_{isj}})}\right) 
\end{eqnarray*}
where $B$ is the total number of MCMC iterations.

\vspace{0.5cm}


\bibliography{SurvReg}
\bibliographystyle{bka}

\end{document}